\documentclass[12pt]{article}
\usepackage{amsmath,cite,epsfig,a4,times,euscript,listings,amssymb}
\usepackage[text-hat-accent]{euler}
\usepackage{graphics}
\usepackage[usenames]{color}
\usepackage{bm}
\renewcommand{\baselinestretch}{1.1}
\newcommand{\myTitle}[1]{\begin{center}{\bf\Huge #1}\\[5ex]\end{center}}
\newcommand{\myAuthor}[1]{\begin{center}{\Large #1}\\[2ex]\end{center}}
\newcommand{\myAffiliation}[1]{\\[1ex]{\it\large #1}}

\newcommand{\myDate}{\begin{center}{\large\today}\\[5ex]\end{center}}
\newcommand{\myAbstract}[1]{\begin{center}\renewcommand{\baselinestretch}{1}{\bf Abstract}\\[2ex]\parbox{0.8\linewidth}{\small\hspace{15pt} #1}\end{center}\vspace{\baselineskip}}
\newcommand{\myReport}[1]{\hspace{\fill} #1}
\newcommand{\myPreprint}[1]{}
\newcommand{\myKeywords}[1]{}

\newcommand{\myScript}[1]{\EuScript{#1}}

\newcommand{\slashp}{p\hspace{-6.5pt}/}

\newcommand{\slashk}{k\hspace{-6.5pt}/}

\newcommand{\Appendix}[1]{Appendix~\ref{#1}}

\newcommand{\Equation}[1]{Eq.~(\ref{#1})}
\newcommand{\ie}{{\it i.e.}}

\newcommand{\eg}{{\it e.g.}}
\newcommand{\Ord}{\myScript{O}}

\newcommand{\Tr}{\mathrm{Tr}}
\newcommand{\imag}{\mathrm{i}}

\newcommand{\vep}{\epsilon}

\newcommand{\gQCD}{g_\mathrm{s}}
\newcommand{\gQED}{e}

\newcommand{\ANG}[1]{\langle#1\rangle}
\newcommand{\SQR}[1]{[#1]}
\newcommand{\ANGSQR}[1]{\langle#1]}
\newcommand{\SQRANG}[1]{[#1\rangle}

\newcommand{\Arght}[1]{|#1\rangle}

\newcommand{\Srght}[1]{|#1]}
\newcommand{\lop}[2]{#1\!\cdot\!#2}
\newcommand{\Nc}{N_\mathrm{c}}
\newcommand{\sgn}{\mathrm{sgn}}
\newcommand{\alphaS}{\alpha_{\mathrm{s}}}

\newcommand{\kapp}{\kappa}
\newcommand{\kstr}{\kappa^*}
\newcommand{\cGamma}{c_{\Gamma}}
\newcommand{\kT}{k_T}
\newcommand{\kTsq}{|k_T^2|}
\newcommand{\RT}{\bigg(\frac{\mu^2}{\kTsq}\bigg)^\vep}
\newcommand{\deltaR}{\delta_{R}}
\newcommand{\nf}{n_f}
\newcommand{\lamlim}{\overset{\Lambda}{\longrightarrow}}
\newcommand{\auxA}{A}
\newcommand{\auxB}{B}
\newcommand{\BLDempty}{\boldsymbol\emptyset}
\newcommand{\BLDto}{\boldsymbol\to}
\newcommand{\BLDg}{\boldsymbol g}
\newcommand{\BLDq}{\boldsymbol q}
\newcommand{\BLDh}{\boldsymbol H}
\newcommand{\BLDqB}{\bar{\boldsymbol q}}
\newcommand{\BLDstar}{\boldsymbol\star}
\newcommand{\BLDgSTR}{\BLDg^{\BLDstar}}
\newcommand{\BLDpls}{\boldsymbol+}
\newcommand{\BLDmin}{\boldsymbol-}
\newcommand{\BLDePLS}{\boldsymbol e^{\BLDpls}}
\newcommand{\BLDeMIN}{\boldsymbol e^{\BLDmin}}

\newcommand{\qq}{q}
\newcommand{\qb}{\bar{q}}
\newcommand{\QQ}{Q}
\newcommand{\QB}{\bar{Q}}
\newcommand{\ePLS}{e^{+}}
\newcommand{\eMIN}{e^{-}}
\newcommand{\tree}{\mathrm{tree}}
\newcommand{\fullAmpTree}{\myScript{M}^{\tree}}
\newcommand{\fullAmp}{\myScript{M}}
\newcommand{\Amp}{\myScript{A}}
\newcommand{\AmpTree}{\Amp^{\tree}}
\newcommand{\Vamp}{\myScript{V}}
\newcommand{\Famp}{\myScript{F}}
\newcommand{\Tamp}{\myScript{T}}
\newcommand{\AmpRt}{\Amp^{\mathrm{Rt}}}
\newcommand{\mysign}{\Upsilon}
\newcommand{\auxq}{\mathrm{aux}\textrm{-}\mathrm{q}}
\newcommand{\auxg}{\mathrm{aux}\textrm{-}\mathrm{g}}
\newcommand{\reLab}{\mathrm{re}}
\newcommand{\NLO}{\mathrm{NLO}}
\newcommand{\procgg}{\emptyset\to g^\star gg}
\newcommand{\procqq}{\emptyset\to g^\star q\bar{q}}
\newcommand{\procH}{\emptyset\to g^\star g H}
\newcommand{\procee}{\emptyset\to g^\star \bar{q}q e^{+} e^{-}}
\newcommand{\procqqgg}{\emptyset\to \bar{q}q gg}
\newcommand{\Cone}{c_1}
\newcommand{\Cthree}{c_3}

\newcommand{\tweakcodepar}[3]%
  {\vspace{#1ex}\newline\noindent\hspace*{4.0ex}{\small\tt #3}\vspace{#2ex}\newline\noindent}

\usepackage{color}
\usepackage[makeroom]{cancel}
\usepackage[normalem]{ulem}
\definecolor{pkcolor}{rgb}{0,0.1,0.7}

\newcommand\pkout{\marginpar{\color{pkcolor}$\clubsuit$}\bgroup\markoverwith{\color{pkcolor}{\rule[0.4ex]{2pt}{0.8pt}}}\ULon}

\definecolor{avhcolor}{rgb}{1.0,0.0,0.0}

\newcommand\avhout{\marginpar{\color{avhcolor}$\clubsuit$}\bgroup\markoverwith{\color{avhcolor}{\rule[0.4ex]{2pt}{0.8pt}}}\ULon}

\begin{document}

\myReport{IFJPAN-IV-2022-20}
\myPreprint{}\\[2ex]

\myTitle{%
One-loop gauge invariant amplitudes\\[0.5ex] with a space-like gluon
}

\myAuthor{%
Etienne~Blanco\footnote{etienne.blanco@ifj.edu.pl},
Alessandro Giachino\footnote{alessandro.giachino@ifj.edu.pl},
Andreas~van~Hameren\footnote{andre.hameren@ifj.edu.pl}
\myAffiliation{%
Institute of Nuclear Physics Polisch Academy of Sciences,\\[-0.5ex]%
PL-31342 Krak\'ow, Poland
}\\[2ex]
Piotr Kotko\footnote{piotr.kotko@fis.agh.edu.pl}
\myAffiliation{%
AGH University Of Science and Technology, Physics Faculty,\\[0.5ex]
S.~Mickiewicza 30, 30-059 Krak\'ow, Poland
}
}

\myDate

\myAbstract{Nowadays the particle physics has entered an era where high precision calculations are required in order to compare the theoretical predictions with the experimental data. In this paper, we explicitly compute the  virtual contributions for  the space-like one-jet processes, $\procgg,\procqq,\procH$ and $\procee$ within the auxiliary parton method. Our results, which are expected to play an important role in high precision description of small $x$ physics, explicitly confirm the conjecture developed in Ref. \cite{vanHameren:2022mtk}, thus helping to bridge the gap between lowest order calculations and NLO corrections within hybrid $k_T$-factorization scheme.
}

\myKeywords{QCD}

%\begin{document}
%

\newpage%
\setcounter{tocdepth}{1}
\tableofcontents

%%%%%%%%%%%%%%%%%%%%%%%%%%%%%%%%%%%%%%%%%%%%%%%%%%%%%%%%%%%%%%%%%%%%%%%%%%%%%%%%
%%% Introduction %%%%%%%%%%%%%%%%%%%%%%%%%%%%%%%%%%%%%%%%%%%%%%%%%%%%%%%%%%%%%%%
%%%%%%%%%%%%%%%%%%%%%%%%%%%%%%%%%%%%%%%%%%%%%%%%%%%%%%%%%%%%%%%%%%%%%%%%%%%%%%%%
\section{Introduction}

The fundamental idea that guides phenomenological investigations in high-energy hadronic physics is factorization.
In collinear factorization (see \eg~\cite{Collins:2011zzd}) it is assumed that the incoming partons carry only longitudinal momenta.
More precisely, the transverse momenta in the hard process are power-suppressed, while in the soft part they are integrated over.
In the small $x$ limit of perturbative QCD,  the centre-of-mass energy $\sqrt{s}$  is much bigger than any hard scales of the problem and 
typically the leading power collinear approximation is not adequate.
Instead, the so-called High Energy Factorization  (HEF),  also called $k_T$-factorization,  applies  \cite{Catani:1990eg,Catani:1990xk}.
The key difference between the $k_T$-factorization and the collinear factorization is that, in  the former, both the hard part and the soft hadronic part depend on the parton transverse momenta $k_T$. This implies that  the  hard matrix elements include explicit higher powers. 
 
In the small $x$ regime, the collinear Parton Distribution Functions (PDFs) are replaced by unintegrated PDFs (with explicit transverse momentum dependence) that follow the BFKL evolution equations  \cite{Kuraev:1976ge,Kuraev:1977fs,Fadin:1975cb,Lipatov:1976zz,Balitsky:1978ic,Lipatov:1985uk,Lipatov:1996ts}.
%\cite{Kuraev:1977fs,Kuraev:1976ge,Fadin:1975cb,Lipatov:1976zz,Balitsky:1978ic,Lipatov:1985uk,Kuraev:1976ge,Lipatov:1976zz,Lipatov:1996ts}. 
When there is an asymmetry in the longitudinal fractions carried by the colliding initial state partons  (a situation present in forward jet production) the HEF can be reduced to the so-called hybrid $k_T$-factorization \cite{Dumitru:2005gt,Deak:2009xt}, where  only one initial-state parton has an explicit dependence on the transverse momentum $k_T$.
This factorization scheme requires partonic hard matrix elements with one space-like (off-shell) initial-state parton. 
Any physically relevant scattering amplitude must be gauge invariant, \ie\ satisfy freedom in the choice of gluon propagators and Ward identity, but these two conditions only hold if all external particles are on-shell. For this reason, the calculation of amplitudes with off-shell intial state cannot be obtained by a naive application of the conventional QCD Feynman rules.
In the literature, this problem was solved at tree-level by the Lipatov's effective action \cite{Lipatov:1995pn,Antonov:2004hh}, and methods which manually restore gauge invariance. The latter are based on Ward identities \cite{vanHameren:2012uj}, matrix elements of straight infinite Wilson lines \cite{Kotko:2014aba}, or  auxiliary parton method  \cite{vanHameren:2012if}, which is the method chosen in the present work.

The idea of the auxiliary parton method relies on embedding the off-shell process into a new process in which the off-shell gluon is replaced by an auxiliary on-shell quark anti-quark pair or by two auxiliary gluons, whose momenta are written as a function of a dimensionless parameter called 
 $\Lambda$ \cite{vanHameren:2012if}. Being this new process on-shell, the constructed amplitude is gauge invariant for any value of $\Lambda$. 
 Most important, in the $\Lambda \to \infty$ limit,  the constructed amplitude is  the desired off-shell scattering amplitude of the original process. 
At Leading Order (LO) one can use equivalently auxiliary quarks or auxiliary gluons, provided to properly take care of the color factors.  Moving to Next-to Leading Order (NLO) inevitably leads to infrared divergences in real and virtual contributions, which do not cancel for processes with colored partons and thus need to be treated by a renormalization of proper operators.
In collinear factorization, this is rigorously established (partly through renormalization of the PDFs) for all orders in perturbation theory \cite{Ellis:1978ty} while, in hybrid $k_T$-factorization, an equivalent rigorous all-order proof has still to be developed. However, working order-by-order, there has been great progress in $k_T$ factorization, both concerning the evolution of unintegrated PDFs and the partonic processes. The NLO BFKL kernel is known \cite{Fadin:1998py,Ciafaloni:1998gs,Kotikov:2000pm} as well as the NLO Balitsky-Kovchegov equation \cite{Balitsky:2007feb} and the whole B-JIMWLK equation  \cite{Balitsky:2013fea,Kovner:2013ona}. In the language of impact factors, several NLO results exist, for example the $\gamma^*\rightarrow \bar{q}q$ impact factor \cite{Bartels:2002uz,Balitsky:2012bs,Beuf:2016wdz,Boussarie:2016ogo,Chachamis:2013bwa}, Higgs plus a hadron \cite{Celiberto:2022fgx}. In Color Glass Condensate theory, the NLO results include the single inclusive jet production \cite{Chirilli:2011km}, dijet production in DIS \cite{Roy:2019hwr,Caucal:2021ent}. In addition, there are NLO calculations in the context of the Lipatov's effective action \cite{Hentschinski:2011tz,Chachamis:2012gh,Chachamis:2012cc,Hentschinski:2014lma,Nefedov:2017qzc,Nefedov:2019mrg,Nefedov:2020ecb,Hentschinski:2020tbi}.

The motivation for the following work is provided by tremendous progress in calculating hard matrix elements in collinear factorization, where, by now, any tree or loop-level process can be computed automatically. 

In $k_T$ factorization, on the other hand, so far only tree-level processes are fully automated. In particular, the KaTie Monte Carlo \cite{vanHameren:2016kkz} allows for efficient computation of any Standard Model processes within the $k_T$ factorization at tree-level, based on the embedding method \cite{vanHameren:2012if}.
The automation at loop-level is still not achieved, but there has been considerable progress. 
In Ref.~\cite{Blanco:2020akb} the embedding method was used to obtain all-plus helicity off-shell gauge invariant amplitudes at one loop. These amplitudes are all finite and it turns out that there is ``auxiliary parton universality'' property, similar to tree-level, namely, one obtains the same result irrespectively whether the auxiliary parton line is a quark line or gluon line. In \cite{vanHameren:2022mtk} the authors made the first rigorous attempt to understand the auxiliary parton approach to $k_T$ factorization at NLO in a general setting.
It was shown that at NLO the auxiliary parton universality is violated, but these violations are process independent.
 Moreover, the authors identified the explicit structure of the virtual and real contributions based on a conjecture regarding the $\Lambda$-dependent virtual contributions.

Let us now summarize the key properties of our approach and outline how we plan to use it in actual cross-section calculations. First, computations of loop diagrams are easier using the auxiliary parton approach than explicitly computing loop diagrams using, for example, the Lipatov's effective action; clearly the approach utilizes existing on-shell one-loop results. Furthermore, although the present work provides analytic results, it seems that the approach can be performed numerically, which would allow for full automation of one-loop computations. Second, the auxiliary parton approach gives in general different results for the divergent parts than the Lipatov's effective action approach. This fact should not be surprising as the auxiliary parton approach uses an entirely different regularization scheme. Full cross-section computation requires also real corrections and proper subtractions of infrared sensitive terms. Since the auxiliary parton approach is already fully understood at tree-level, it is possible to recognize all singular regions with auxiliary partons and construct a general framework for computing the real corrections, in the spirit of the dipole subtraction method \cite{Catani:1996vz} or the antenna method \cite{Kosower:1997zr,Kosower:2003bh}. Once this is done, proper infrared subtraction terms, consistent with the $k_T$ factorization can be constructed. Currently, there is an ongoing research on the real corrections within the auxiliary parton approach. Finally, it is important to stress that the ultimate goal is to build a Monte Carlo code that can handle NLO computation of a wide range of observables within the $k_T$ factorization approach and the auxiliary parton approach seems well suited for that purpose.

In this paper, we explicitly calculate the virtual contributions for the following processes: $\procgg,\procqq,\procH$ and $\procee$, with explicit helicity projections using the auxiliary parton approach.
In each case we check their collinear limit, \ie\ the limit when the gluon has vanishing transverse momentum and becomes on-shell. Our results explicitly confirm the conjecture developed in \cite{vanHameren:2022mtk} regarding the structure of the divergent auxiliary parton depended terms.
Originally, the auxiliary parton method was introduced just as a trick to obtain LO helicity amplitudes involving space-like partons and there were not any a priori theoretical justifications which guarantee its application to compute NLO cross sections. Our explicit calculation of the virtual contributions within the auxiliary parton method confirms for the considered processes that the auxiliary parton universality is violated, but in a universal way, thus opening the way for the establishment of a rigorous formalism to calculate higher order corrections suitable for a numerical approach.

%%%%%%%%%%%%%%%%%%%%%%%%%%%%%%%%%%%%%%%%%%%%%%%%%%%%%%%%%%%%%%%%%%%%%%%%%%%%%%%%
%%% Definitions %%%%%%%%%%%%%%%%%%%%%%%%%%%%%%%%%%%%%%%%%%%%%%%%%%%%%%%%%%%%%%%%
%%%%%%%%%%%%%%%%%%%%%%%%%%%%%%%%%%%%%%%%%%%%%%%%%%%%%%%%%%%%%%%%%%%%%%%%%%%%%%%%
\section{The auxiliary parton method}
\subsection{Definitions}
Our results are written in the language of the spinor helicity method (see Appendix \ref{weylspinors}) and $SU(N_c)$ color decomposition.
When presenting helicity amplitudes, it is most convenient to imagine all momenta to be outgoing, and have momentum conservation as
%
%%%%%%%%%%%%%%%%%%%%%%%%%%%%%%%%%%%%%%%%
\begin{equation}
\sum_{i=1}^{n}k_i^\mu = 0
~,
\end{equation}
%%%%%%%%%%%%%%%%%%%%%%%%%%%%%%%%%%%%%%%%
%
where $k_i^\mu$ for $i=1,\ldots,n$ are the external momenta.
This implies that the initial-state momenta have negative energy.
The space-like initial-state momentum will always be the first one.
It has an explicit longitudinal component and transverse components, and we find it most convenient to write it as
%
%%%%%%%%%%%%%%%%%%%%%%%%%%%%%%%%%%%%%%%%
\begin{equation}
k_1^\mu = k^\mu = p^\mu + \kT^\mu
~,
\label{eq:kT_offshell}
\end{equation}
%%%%%%%%%%%%%%%%%%%%%%%%%%%%%%%%%%%%%%%%
%
where $p^\mu$ is the light-like longitudinal component that will appear in spinor products etc., and $\lop{p}{\kT}=0$.
It is eventually equal to
%
%%%%%%%%%%%%%%%%%%%%%%%%%%%%%%%%%%%%%%%%
\begin{equation}
p^\mu = -xP_{\mathrm{hadron}}^\mu
~,
\end{equation}
%%%%%%%%%%%%%%%%%%%%%%%%%%%%%%%%%%%%%%%%
%
where $P_{\mathrm{hadron}}^\mu$ is the (positive-energy) hadron momentum and $x$ its positive momentum fraction.
The definition \eqref{eq:kT_offshell} is actually consistent with the
so-called quasi Regge kinematics in high energy scattering, see eg. \cite{Antonov:2004hh}.
Since $k^\mu$ is space-like, its square is negative, which we express with absolute value symbols as follows:
%
%%%%%%%%%%%%%%%%%%%%%%%%%%%%%%%%%%%%%%%%
\begin{equation}
k^2 = -\kTsq
~.
\end{equation}
%%%%%%%%%%%%%%%%%%%%%%%%%%%%%%%%%%%%%%%%

In order to extract helicity amplitudes with the auxiliary parton method, we would like to apply exact kinematics, and introduce auxiliary parton momenta
% 
%%%%%%%%%%%%%%%%%%%%%%%%%%%%%%%%%%%%%%%%
\begin{align}
p_{\auxA}^\mu &= \hspace{5.4ex}\Lambda p^\mu \hspace{0ex}+ \alpha q^\mu + \hspace{5.5ex}\beta \kT^\mu
\\
p_{\auxB}^\mu &= (1-\Lambda) p^\mu - \alpha q^\mu + (1-\beta) \kT^\mu
~,
\end{align}
%%%%%%%%%%%%%%%%%%%%%%%%%%%%%%%%%%%%%%%%
%
where $q^\mu$ is light-like with $p\!\cdot\!q>0$ and $q\!\cdot\!\kT=0$, and where
%
%%%%%%%%%%%%%%%%%%%%%%%%%%%%%%%%%%%%%%%%
\begin{equation}
\alpha = \frac{-\beta^2\kT^2}{\Lambda(p+q)^2}
\quad,\quad
\beta = \frac{1}{1+\sqrt{1-1/\Lambda}}
\quad.
\label{auxmom2}
\end{equation}
%%%%%%%%%%%%%%%%%%%%%%%%%%%%%%%%%%%%%%%%
%
With this choice, the momenta $p_{\auxA}^\mu,p_{\auxB}^\mu$ satisfy the relations
%
%%%%%%%%%%%%%%%%%%%%%%%%%%%%%%%%%%%%%%%%
\begin{equation}
p_{\auxA}^2 = p_{\auxB}^2 = 0
\quad,\quad
p_{\auxA}^\mu+p_{\auxB}^\mu = p^\mu+\kT^\mu
\end{equation}
%%%%%%%%%%%%%%%%%%%%%%%%%%%%%%%%%%%%%%%%
%
for any value of the parameter $\Lambda$, which is why we call it ``exact''.
Thus, the momenta $p_A$ and $p_B$ represent the momenta of external auxiliary partons (a $q$-$\overline{q}$ pair or a pair of gluons) whose sum gives the momentum of the off-shell gluon \eqref{eq:kT_offshell}. We now consider an actual amplitude with these external partons and \emph{define} the gauge invariant amplitude with a space-like gluon as follows:
%
%Amplitudes with a space-like gluon are obtained from amplitudes with an auxiliary quark pair following
%
%%%%%%%%%%%%%%%%%%%%%%%%%%%%%%%%%%%%%%%%
\begin{equation}
\frac{1}{\Lambda}
\,\fullAmp\Big(q\big(p_{\auxA}\big),\bar{q}\big(p_{\auxB}\big),\ldots\Big)
  \;\overset{\Lambda\to\infty}{\longrightarrow}\;
\fullAmp\big(g^\star(p+\kT),\ldots\big)
~,
\label{lamlim}
\end{equation}
%%%%%%%%%%%%%%%%%%%%%%%%%%%%%%%%%%%%%%%%
%
where the ellipses stand for other particles and partons involved in the scattering process. Above, we assumed that the auxiliary pair are quarks.
In earlier work, \eg~\cite{Blanco:2020akb}, there is a factor $x|\kT|/\gQCD$ included on the left-hand side.
The factor $x$ is included implicitly here because we chose to parametrize the auxiliary parton momenta in terms of $p$ instead of $P_{\mathrm{hadron}}$.
The factor $|\kT|$ is omitted to give the amplitudes presented here a more natural analytic form.
We also decided to omit an imaginary unit from the amplitudes for convenience, and we summarize with the statement that
%
%%%%%%%%%%%%%%%%%%%%%%%%%%%%%%%%%%%%%%%%
\begin{equation}
\textit{\parbox{76ex}{a factor $\imag|\kT|$ is omitted from the presented amplitudes and must be included in the end.}}
\nonumber
\end{equation}
%%%%%%%%%%%%%%%%%%%%%%%%%%%%%%%%%%%%%%%%
%
%\color{red}
The coupling $\gQCD$ will be considered to be equal to $1$ in this write-up.
We must just imagine that in the end, tree-level amplitudes get a factor $\gQCD^{n-2}$, and one-loop amplitudes get a factor $\gQCD^{n}$, where $n$ is the number of partons involved in the scattering process.
The procedure in \Equation{lamlim} of taking $\Lambda\to\infty$ on auxiliary parton momenta and dividing by $\Lambda$ will also be abbreviated with
%
%%%%%%%%%%%%%%%%%%%%%%%%%%%%%%%%%%%%%%%%
\begin{equation}
\fullAmp\big(\auxA,\auxB,\ldots\big)
  \;\overset{\Lambda}{\longrightarrow}\;
\fullAmp\big(g^\star,\ldots\big)
~.
\label{LargeL}
\end{equation}
%%%%%%%%%%%%%%%%%%%%%%%%%%%%%%%%%%%%%%%%

The transverse momentum $\kT^\mu$ can be written in terms of polarization vectors constructed with the help of $q^\mu$ 
%
%%%%%%%%%%%%%%%%%%%%%%%%%%%%%%%%%%%%%%%%
\begin{equation}
\kT^\mu = -\frac{\kstr}{\sqrt{2}}\,\frac{\ANGSQR{q|\gamma^\mu|p}}{\sqrt{2}\,\ANG{qp}}
          -\frac{\kapp}{\sqrt{2}}\,\frac{\ANGSQR{p|\gamma^\mu|q}}{\sqrt{2}\,\SQR{pq}}
\quad\textrm{with}\quad
\kapp = \frac{\ANGSQR{q|\slashk|p}}{\ANG{qp}}
\quad,\quad
\kstr = \frac{\ANGSQR{p|\slashk|q}}{\SQR{pq}}
~.
\end{equation}
%%%%%%%%%%%%%%%%%%%%%%%%%%%%%%%%%%%%%%%%
%
Both $\kapp$ and $\kstr$ do not implicitly depend on $x$, since both $\Arght{p}$ and $\Srght{p}$ scale as $\sqrt{x}$ for $x>0$.
Also, they do not depend on $q^\mu$, and we have
%
%%%%%%%%%%%%%%%%%%%%%%%%%%%%%%%%%%%%%%%%
\begin{equation}
\kapp\kstr = \kTsq
~.
\end{equation}
%%%%%%%%%%%%%%%%%%%%%%%%%%%%%%%%%%%%%%%%
%
Realizing that $\sqrt{1-1/\Lambda}=1/\beta-1$, 
the Weyl spinors for $p_{\auxA}^\mu,p_{\auxB}^\mu$ can be written  as
%
%%%%%%%%%%%%%%%%%%%%%%%%%%%%%%%%%%%%%%%%
\begin{align}
\Arght{\auxA} = \sqrt{\Lambda}\,\Arght{p}
            - \frac{\beta\kstr}{\sqrt{\Lambda}\,\ANG{qp}}\,\Arght{q}
\quad&,\quad
\Srght{\auxA} = \sqrt{\Lambda}\,\Srght{p}
            - \frac{\beta\kapp}{\sqrt{\Lambda}\,\SQR{pq}}\,\Srght{q}
\quad,
\label{AAprimeSpinors}\\
\Arght{\auxB} = \sqrt{\Lambda-1}\,\Arght{p}
            + \frac{\beta\kstr}{\sqrt{\Lambda}\,\ANG{qp}}\,\Arght{q}
\quad&,\quad
\Srght{\auxB} = -\sqrt{\Lambda-1}\,\Srght{p}
            - \frac{\beta\kapp}{\sqrt{\Lambda}\,\SQR{pq}}\,\Srght{q}
\quad.
\notag
\end{align}
%%%%%%%%%%%%%%%%%%%%%%%%%%%%%%%%%%%%%%%% 
%
Explicitly, the calculation of  $\Lambda\to\infty$ limit of the left side of \Equation{LargeL} consists in performing the substitutions reported in Eq.  \ref{AAprimeSpinors},  for example $\ANG{\auxA4}\to\sqrt{\Lambda}\ANG{p4}$, in a precise order respecting the following sequence:
%
%%%%%%%%%%%%%%%%%%%%%%%%%%%%%%%%%%%%%%%%
\begin{itemize}
\item{1)}
\begin{equation}
\ANG{\auxA\auxB} \to -\kstr
\quad,\quad
\SQR{\auxA\auxB} \to -\kapp
\quad,\quad
p_{\auxA}^\mu+p_{\auxB}^\mu \to k^\mu
~.
\end{equation}
%%%%%%%%%%%%%%%%%%%%%%%%%%%%%%%%%%%%%%%%
%
The latter implies also 
%%%%%%%%%%%%%%%%%%%%%%%%%%%%%%%%%%%%%%%%
\begin{equation}
s_{\auxA\auxB}=(p_{\auxA}+p_{\auxB})^2\to k^2=-\kapp\kstr
\quad,\quad
t_{\auxA\auxB i}=(p_{\auxA}+p_{\auxB}+p_i)^2\to(k+p_i)^2=s_{ki}
\end{equation}
%%%%%%%%%%%%%%%%%%%%%%%%%%%%%%%%%%%%%%%%
%
etc.
\item 2)
%%%%%%%%%%%%%%%%%%%%%%%%%%%%%%%%%%%%%%%%
\begin{equation}
p_{\auxA}^\mu \to \Lambda p^\mu %+ \srac{1}{2}\kT^\mu
\quad,\quad
p_{\auxB}^\mu \to -\Lambda p^\mu %+ \srac{1}{2}\kT^\mu - xp^\mu
\end{equation}
%%%%%%%%%%%%%%%%%%%%%%%%%%%%%%%%%%%%%%%%
%
so that
%
%%%%%%%%%%%%%%%%%%%%%%%%%%%%%%%%%%%%%%%%
\begin{equation}
s_{\auxA i}\to\Lambda s_{pi}
\quad,\quad
s_{\auxB i}\to-\Lambda s_{pi}
~,
\end{equation}
%%%%%%%%%%%%%%%%%%%%%%%%%%%%%%%%%%%%%%%%
%
\item 3)
%
%%%%%%%%%%%%%%%%%%%%%%%%%%%%%%%%%%%%%%%%
\begin{equation}
\Arght{\auxA}\to\sqrt{\Lambda}\,\Arght{p}
\quad,\quad
\Srght{\auxA}\to \sqrt{\Lambda}\,\Srght{p}
\quad,\quad
\Arght{\auxB}\to\sqrt{\Lambda}\,\Arght{p}
\quad,\quad
\Srght{\auxB}\to-\sqrt{\Lambda}\,\Srght{p}
~.
\end{equation}
%%%%%%%%%%%%%%%%%%%%%%%%%%%%%%%%%%%%%%%%
%
\end{itemize}
These operations give the correct result under the condition that all terms in an expression to which the operations are applied exhibit at most the leading power behavior of $\Lambda^p$ with $p=1$.
We encountered only one expression with higher-order terms that cancel among each other and require including the $1/\sqrt{\Lambda}$ terms in \Equation{AAprimeSpinors}.
It will be mentioned in \Appendix{App:eeauxg}.

We follow the conventions in literature, in particular~\cite{Bern:1990ux,Bern:1993mq,Bern:1994fz,Bern:1997sc}, to decompose amplitudes into partial amplitudes.
Consider for example multi-gluon amplitudes.
At tree-level, we have
%
%%%%%%%%%%%%%%%%%%%%%%%%%%%%%%%%%%%%%%%%
\begin{equation}
\fullAmpTree(1,\ldots,n) = \sum_{\sigma\in S_n/Z_n}\Tr\big(T^{a_{\sigma(1)}}\cdots T^{a_{\sigma(n)}}\big)\AmpTree(\sigma(1),\ldots,\sigma(n))
\label{Eq:def09}
~.
\end{equation}
%%%%%%%%%%%%%%%%%%%%%%%%%%%%%%%%%%%%%%%%
%
The arguments of the amplitude on the left-hand side indicate that it depends on $n$ gluon momenta, helicities, and colors.
We note, that it is common to include an index $n$ for an amplitude, but to make our notation simpler we shall omit that index, as the number of external states is implied by the argument list.
%%%%%%%%%%%%%%%%%%%%%%%%%%%%%%%%%%%%%%%%
%\begin{equation}
%\textrm{\parbox{76ex}{We will omit a subscript $n$ for such functions, %since it is implied by the argument list.}}
%\nonumber
%\end{equation}
%%%%%%%%%%%%%%%%%%%%%%%%%%%%%%%%%%%%%%%%
%
On the right-hand side, only the color variables $a_i$ are made explicit, elucidating the color decomposition.
The partial amplitude $\AmpTree$ does not depend on these variables.
It is however not symmetric in its arguments enumerating the gluon momenta and helicities.
The sum is over all cyclically nonequivalent  permutations.
Regarding the color algebra, we follow the conventions with
%
%%%%%%%%%%%%%%%%%%%%%%%%%%%%%%%%%%%%%%%%
\begin{equation}
\Tr\big(T^aT^b\big) = \delta^{ab}
\quad,\quad
[T^a,T^b] = \sqrt{2}\,\imag f^{abc}T^c
~,
\end{equation}
%%%%%%%%%%%%%%%%%%%%%%%%%%%%%%%%%%%%%%%%
%
and $f^{123}=1$.
Tree-level amplitudes will carry the superscript as above, while amplitudes without this superscript will always be implied to be at one loop.
The color decomposition for multi-gluon amplitudes becomes
%
%%%%%%%%%%%%%%%%%%%%%%%%%%%%%%%%%%%%%%%%
\begin{multline}
\fullAmp(1,\ldots,n)
= \sum_{\sigma\in S_n/Z_n}
  \Nc\Tr\big(T^{a_{\sigma(1)}}\cdots T^{a_{\sigma(n)}}\big)
  \Amp_1(\sigma(1),\ldots,\sigma(n))
\\
+ \sum_{c=2}^{\lfloor n/2\rfloor+1}\sum_{\sigma\in S_{n;c}/Z_n}
  \Tr\big(T^{a_{\sigma(1)}}\cdots T^{a_{\sigma(c-1)}}\big)
  \Tr\big(T^{a_{\sigma(c)}}\cdots T^{a_{\sigma(n)}}\big)
  \Amp_c(\sigma(1),\ldots,\sigma(n))
~.
\label{Eq:def08}
\end{multline}
%%%%%%%%%%%%%%%%%%%%%%%%%%%%%%%%%%%%%%%%
%
More color structures appear, and $S_{n;c}$ is the subset of $S_{n}$ that leaves the corresponding double trace structure invariant.
%
%%%%%%%%%%%%%%%%%%%%%%%%%%%%%%%%%%%%%%%%
\begin{equation}
\textit{\parbox{76ex}{The subscript $c$ on a partial one-loop amplitude $\Amp_c$ indicates to which color structure it belongs, where $c=1$ always refers the ``leading-color'' tree-level structure.}}
\nonumber
\end{equation}
%%%%%%%%%%%%%%%%%%%%%%%%%%%%%%%%%%%%%%%%
%
The one-loop partial amplitudes are typically decomposed further into more basic building blocks called primitive amplitudes.
They will be labelled with superscripts.
Just like all amplitudes mentioned so far, they are gauge invariant.

The color decomposition examples above are for on-shell gluons, but also valid in case there is a space-like one.
As mentioned earlier, the space-like one will always be labelled as number $1$.
Amplitudes involving the space-like gluons will not carry any special label, but will be implied to be as such by having the argument $1^\star$.
So for example in $\fullAmp(1,2,3,4)$ all gluons are on-shell, while in $\fullAmp(1^\star,2,3,4)$ number $1$ is space-like.
Helicities are indicated with superscripts to the arguments, while the space-like gluon does not come with helicities and will always have superscript $\star$ as in $\AmpTree(1^\star,2^+,3^-,4^-)$.
If arguments do not refer to gluons, then this is indicated with subscripts.
For example, $\fullAmp(1^\star,2,3_{\qb},4_{\qq})$ refers to the process with a space-like gluon, an on-shell gluon, and a quark-antiquark pair.

One particular ``primitive'' amplitude that will appear for many processes with a space-like gluon in the following and deserves a dedicated symbol is
%
%%%%%%%%%%%%%%%%%%%%%%%%%%%%%%%%%%%%%%%%
\begin{equation}
\AmpRt = \cGamma\RT\frac{1}{\vep}\,\Amp^{\tree}
\label{Eq:def06}
~,
\end{equation}
%%%%%%%%%%%%%%%%%%%%%%%%%%%%%%%%%%%%%%%%
%
giving us the opportunity to introduce the one-loop constant
%
%%%%%%%%%%%%%%%%%%%%%%%%%%%%%%%%%%%%%%%%
\begin{equation}
\cGamma = \frac{1}{(4\pi)^{2-\vep}}\frac{\Gamma(1+\vep)\Gamma^2(1-\vep)}{\Gamma(1-2\vep)}
~,
\end{equation}
%%%%%%%%%%%%%%%%%%%%%%%%%%%%%%%%%%%%%%%%
%
the dimensional regularization parameter
%
%%%%%%%%%%%%%%%%%%%%%%%%%%%%%%%%%%%%%%%%
\begin{equation}
\vep = \frac{4-\mathrm{dim}}{2}
~,
\end{equation}
%%%%%%%%%%%%%%%%%%%%%%%%%%%%%%%%%%%%%%%%
%
and the renormalization scale $\mu$.
The amplitude $\AmpRt$ will typically be accompanied by one of the functions
%
%
%%%%%%%%%%%%%%%%%%%%%%%%%%%%%%%%%%%%%%%%
\begin{equation}
\Theta(s) = 2\,\imag\,\pi\,\theta(s)
\quad,\quad
\mysign(s) = \imag\,\pi\,\sgn(s)
\label{Eq:def07}
~,
\end{equation}
%%%%%%%%%%%%%%%%%%%%%%%%%%%%%%%%%%%%%%%%
%
where $s$ is one of the invariants in the process.
Amplitudes coming with this function represent imaginary parts that can be ignored at NLO.

In order to make the correspondence with real-radiation contributions in NLO calculations, it is useful to mention the relation
%
%%%%%%%%%%%%%%%%%%%%%%%%%%%%%%%%%%%%%%%%
\begin{equation}
\gQCD^2\cGamma = \frac{1}{2}\cdot\frac{\alphaS}{2\pi}
                 \frac{(4\pi)^\vep}{\Gamma(1-\vep)} + \Ord\big(\vep^3\big)
~,
\end{equation}
%%%%%%%%%%%%%%%%%%%%%%%%%%%%%%%%%%%%%%%%
%
where the explicit factor $1/2$ is compensated by the factor $2$ in the virtual contribution being $2\mathrm{Re}\big(\fullAmp^{\tree\dagger}\fullAmp\big)$.

\subsection{Analytic continuation}
Our starting point will be expressions from literature that are valid when all invariants are negative, and are analytically continued by imagining those invariants have a small positive imaginary part, \eg
%
%%%%%%%%%%%%%%%%%%%%%%%%%%%%%%%%%%%%%%%%
\begin{equation}
\ln\bigg(\frac{-s_1}{-s_2}\bigg) = \ln(-s_1-\imag\eta) - \ln(-s_2-\imag\eta)
\quad,\quad
\eta>0~.
\end{equation}
%%%%%%%%%%%%%%%%%%%%%%%%%%%%%%%%%%%%%%%%
%
In the auxiliary parton method outlined before, this imaginary part of invariants involving parton $\auxB$ may flip sign, that is
%
%%%%%%%%%%%%%%%%%%%%%%%%%%%%%%%%%%%%%%%%
\begin{equation}
s_{\auxB i} + \imag\eta \to -\Lambda s_{pi} + \imag\eta = -\Lambda(s_{pi}-\imag\eta/\Lambda)
~.
\end{equation}
%%%%%%%%%%%%%%%%%%%%%%%%%%%%%%%%%%%%%%%%
%
This will only be relevant for arguments of the $\ln$ function, and we will present expressions brought to the standard form, with positive imaginary parts for all invariants, using the relation
%
%%%%%%%%%%%%%%%%%%%%%%%%%%%%%%%%%%%%%%%%
\begin{equation}
\ln(s-\imag\eta) = \ln(-s-\imag\eta) + \mysign(s)
~,
\label{Eq:def10}
\end{equation}
%%%%%%%%%%%%%%%%%%%%%%%%%%%%%%%%%%%%%%%%
%
with $\mysign$ defined in \Equation{Eq:def07}.

We also mention here that ratios of invariants raised to a power of the parameter $\vep$ from dimensional regularization (like in \Equation{Eq:def06}) must always be imagined to be expanded in that parameter as
%
%%%%%%%%%%%%%%%%%%%%%%%%%%%%%%%%%%%%%%%%
\begin{equation}
\frac{x^\vep}{\vep^2} = \frac{1}{\vep^2} + \frac{\ln x}{\vep} + \frac{\ln^2x}{2} + \Ord(\vep)
\;\;,\quad
\frac{x^\vep}{\vep} = \frac{1}{\vep} + \ln x + \Ord(\vep)
\;\;,\quad
x^\vep = 1 + \Ord(\vep)
\;\;,
\end{equation}
%%%%%%%%%%%%%%%%%%%%%%%%%%%%%%%%%%%%%%%%
%
where we ignore $\Ord(\vep)$.

\subsection{Auxiliary parton type}
Instead of amplitudes with an auxiliary quark pair, one can also use amplitudes with an auxiliary gluon pair to extract the amplitude with a space-like gluon.
In Appendix~B of~\cite{vanHameren:2022mtk} it is explained that the amplitudes obtained follow the relations
%
%%%%%%%%%%%%%%%%%%%%%%%%%%%%%%%%%%%%%%%%
\begin{align}
\fullAmp^{\tree,\imath_{\auxB}}_{\hspace{2.5ex}i_{\auxA}}(\auxB_{\bar{q}},\auxA_q,\ldots)
&\lamlim
\big(T^{a_1}\big)^{\imath_{\auxB}}_{i_{\auxA}}\fullAmp^{\tree,a_1}(1^\star,\ldots)
~,\\
\fullAmp^{\tree,a_{\auxB}a_{\auxA}}(\auxB,\auxA,\ldots)
&\lamlim
\sqrt{2}\,\imag\,f^{a_{\auxB}a_{\auxA}a_1}\fullAmp^{\tree,a_1}(1^\star,\ldots)
~,\label{Eq:def05}
~.
\end{align}
%%%%%%%%%%%%%%%%%%%%%%%%%%%%%%%%%%%%%%%%
%
Here we made only relevant color indices explicit.
Also realize that in~\cite{vanHameren:2022mtk} the $T$ matrices are normalized differently.
The relations can be inverted as
%
%%%%%%%%%%%%%%%%%%%%%%%%%%%%%%%%%%%%%%%%
\begin{align}
\big(T^{a_1}\big)_{\imath_{\auxB}}^{i_{\auxA}}\fullAmp^{\tree,\imath_{\auxB}}_{\hspace{2.5ex}i_{\auxA}}(\auxB_{\bar{q}},\auxA_q,\ldots)
&\lamlim
\fullAmp^{\tree,a_1}(1^\star,\ldots)
~,\label{Eq:def04}\\
\frac{-\imag}{\sqrt{2}\Nc}\,f^{a_{\auxB}a_{\auxA}a_1}\fullAmp^{\tree,a_{\auxB}a_{\auxA}}(\auxB,\auxA,\ldots)
&\lamlim
\fullAmp^{\tree,a_1}(1^\star,\ldots)
~.
\end{align}
%%%%%%%%%%%%%%%%%%%%%%%%%%%%%%%%%%%%%%%%
%
At one loop, the color structure is richer than at tree-level, and the relations generally only hold for the leading-color, tree-like, color structures.
However, in an NLO calculation, the one-loop structures are contracted with tree-level structures, and we will see in the examples here that only the leading-color structures survive.
%
%
%%%%%%%%%%%%%%%%%%%%%%%%%%%%%%%%%%%%%%%%
\begin{equation}
\textit{\parbox{76ex}{We denote by $\fullAmp^{\NLO}$ those parts of the one-loop amplitude that are relevant at NLO.}}
\nonumber
%\quad\textrm{We denote by $\fullAmp^{\NLO}$ those parts of the one-loop amplitude that are relevant at NLO.}
\end{equation}
%%%%%%%%%%%%%%%%%%%%%%%%%%%%%%%%%%%%%%%%
%

%{\color{red}%
At tree-level, the limit of \Equation{lamlim} is well-defined, but at one loop, it produces terms proportional to $\ln\Lambda$.
Furthermore, the result depends on the type of auxiliary partons.
It involves the leading-color structures, and the form of the dependence was conjectured in \cite{vanHameren:2022mtk} to be
%
%%%%%%%%%%%%%%%%%%%%%%%%%%%%%%%%%%%%%%%%
\begin{align}
\Amp_{1}^{\auxg} &= \Amp_{1} + \cGamma\AmpTree\RT\bigg[
         \frac{1}{\vep}\big(2\ln\Lambda-\imag\pi\big)
        -\frac{1}{\vep^2} + \frac{\pi^2}{3} \bigg]
\label{oneloopauxg}
~,\\
\Amp_{1}^{\auxq} &= \Amp_{1} + \cGamma\AmpTree\RT\bigg[
         \frac{1}{\vep}\big(2\ln\Lambda-\imag\pi\big)
        +\frac{1}{\vep}\frac{13}{6} + \frac{\pi^2}{3} + \frac{83-3\deltaR}{18}
\notag\\&\hspace{28ex}
        +\frac{1}{\Nc^2}\bigg(\frac{1}{\vep^2}+\frac{3}{2\vep}+\frac{7+\deltaR}{2}\bigg)
        -\frac{\nf}{\Nc}\bigg(\frac{2}{3\vep}+\frac{10}{9}\bigg)
        \bigg]
\label{oneloopauxq}
~.
\end{align}
%%%%%%%%%%%%%%%%%%%%%%%%%%%%%%%%%%%%%%%%
%
The conjecture states that $\Amp_{1}$ indeed will be independent of the auxiliary parton type, and produces the expected on-shell limit for $|\kT|\to0$.
More specifically, the divergent part of $\Amp_{1}$ has the on-shell form, but with the longitudinal momentum component $p^\mu$ instead of the momentum $k^\mu$, \eg\ involves the invariants $s_{pi}$ instead of $s_{ki}$.

In the following, we directly present the expressions for $\Amp_{1}$.
Amplitudes coming with sub-leading color structures $\Amp_{i>1}$ do not depend on the type of auxiliary parton or on $\Lambda$.
%
%%%%%%%%%%%%%%%%%%%%%%%%%%%%%%%%%%%%%%%%
\begin{equation}
\textit{\parbox{76ex}{The subtractions of \Equation{oneloopauxg} and \Equation{oneloopauxq} to arrive at $\Amp_1$ will be implied to be included in the operation ``$\lamlim$'' of \Equation{LargeL}.}}
\nonumber
\end{equation}
%%%%%%%%%%%%%%%%%%%%%%%%%%%%%%%%%%%%%%%%
%
Finally, the amplitudes we present are {\em not} UV-subtracted, and include the parameter $\delta_R$ to distinguish between the 't~Hooft-Veltman scheme of dimensional regularization ($\delta_R=1$) and the four-dimensional helicity scheme ($\delta_R=0$).

\subsection{Symmetrization}
We find that, despite the subtractions of \Equation{oneloopauxg} and \Equation{oneloopauxq}, 
amplitudes with extra quark pair still depend on the type of auxiliary partons.
The dependence is caused by terms proportional to $\mysign(s)$ originating from \Equation{Eq:def10} for some invariants.
We argue that the difference vanishes upon summation over helicities and symmetrization as follows.

In~\cite{vanHameren:2022mtk} it was argued that the auxiliary parton method must be considered to be applied at the cross-section level, rather than just at the amplitude level, in order to take the real radiation contribution into account correctly.
This means that also the auxiliary partons must be summed over their helicities before taking the $\Lambda$ limit.
At tree-level, already for the amplitudes, only the opposite-helicity configurations for the auxiliary partons contribute, and either configuration gives the same result.
Consider auxiliary quarks.
For the virtual contribution we can then write
%
%%%%%%%%%%%%%%%%%%%%%%%%%%%%%%%%%%%%%%%%
\begin{align}
\EuScript{V}_{\qq} &= 2\mathrm{Re}\Big\{
   \fullAmp^{\tree\,\dagger}(\auxA_{\qb}^+,\auxB_{\qq}^-)
   \fullAmp(\auxA_{\qb}^+,\auxB_{\qq}^-)
  +\fullAmp^{\tree\,\dagger}(\auxA_{\qb}^-,\auxB_{\qq}^+)
   \fullAmp(\auxA_{\qb}^-,\auxB_{\qq}^+)\Big\}
\notag\\&=
2\mathrm{Re}\Big\{\fullAmp^{\tree\,\dagger}(\auxA_{\qb}^+,\auxB_{\qq}^-)\big[
    \fullAmp(\auxA_{\qb}^+,\auxB_{\qq}^-) 
  + \fullAmp(\auxA_{\qb}^-,\auxB_{\qq}^+)\big]\Big\}
~.
\label{Eq:virthelsum}
\end{align}
%%%%%%%%%%%%%%%%%%%%%%%%%%%%%%%%%%%%%%%%
%
Here we used the ``$\dagger$'' to indicate complex conjugation, and only wrote the helicity sum for the auxiliary partons explicitly.
Considering the definition of the momenta $p_{\auxA},p_{\auxB}$, the anti-quark in the amplitude must be seen as an initial-state quark, and one could argue that the contribution from anti-quark scattering must be included too.
For the amplitudes, this simply means that the role of $\auxA,\auxB$ must be reversed.
At tree-level, this just amounts to an overall minus sign, and we can write
%
%%%%%%%%%%%%%%%%%%%%%%%%%%%%%%%%%%%%%%%%
\begin{align}
\EuScript{V}_{\qq}+ \EuScript{V}_{\qb} &= 2\mathrm{Re}\Big\{
  \fullAmp^{\tree\,\dagger}(\auxA_{\qb}^+,\auxB_{\qq}^-)\big[
      \fullAmp(\auxA_{\qb}^+,\auxB_{\qq}^-) 
    + \fullAmp(\auxA_{\qb}^-,\auxB_{\qq}^+)
\notag\\&\hspace{24ex}
    - \fullAmp(\auxB_{\qb}^+,\auxA_{\qq}^-) 
    - \fullAmp(\auxB_{\qb}^-,\auxA_{\qq}^+)\big]\Big\}
~.
\label{symauxq}
\end{align}
%%%%%%%%%%%%%%%%%%%%%%%%%%%%%%%%%%%%%%%%
%
We will see that in this combination of one-loop amplitudes, the problematic terms that cause the auxiliary parton dependence drop out.

Regarding auxiliary gluons, one can argue that the same symmetrization regarding the role of $\auxA,\auxB$ must be applied.
%
%In practical cross section calculations involving one initial-state gluon and say $n$ gluons, one only symmetrizes over the final-state gluons by including a factor $1/n!$, because the phase space separation between initial-state and final-state is trivial.
%
%In a bigger scope one could say that every gluon should have a \PK{chance} to be the initial-state one, and then the symmetry factor should be $1/(n+1)!$.
%
%This then implies that one should symmetrize over the role of the auxiliary gluons, and take the combination
In a bigger scope, one could say that the hard process is just a factorized  part of the full process, the latter including the diagrams with reversed roles. Thus, following the same procedure as for auxiliary quarks, we should take the combination
%%%%%%%%%%%%%%%%%%%%%%%%%%%%%%%%%%%%%%%%
\begin{equation}
      \fullAmp(\auxA^+,\auxB^-) 
    + \fullAmp(\auxA^-,\auxB^+)
    - \fullAmp(\auxB^+,\auxA^-) 
    - \fullAmp(\auxB^-,\auxA^+)
~,
\label{symauxg}
\end{equation}
%%%%%%%%%%%%%%%%%%%%%%%%%%%%%%%%%%%%%%%%
%
and include a factor $1/2$.
At tree-level, this does not make any difference.
We will see that at one loop this causes the problematic terms to drop out again.
%}%\color{red}

%%%%%%%%%%%%%%%%%%%%%%%%%%%%%%%%%%%%%%%%%%%%%%%%%%%%%%%%%%%%%%%%%%%%%%%%%%%%%%%%
%%% 2->1 processes %%%%%%%%%%%%%%%%%%%%%%%%%%%%%%%%%%%%%%%%%%%%%%%%%%%%%%%%%%%%%
%%%%%%%%%%%%%%%%%%%%%%%%%%%%%%%%%%%%%%%%%%%%%%%%%%%%%%%%%%%%%%%%%%%%%%%%%%%%%%%%
\section{Results}
In this section we present our results for the processes, $\procgg,\procqq,\procH$ and $\procee$,  using the auxiliary parton method described above.
All of them carry one power of the strong coupling constant, but are increasingly less trivial regarding the kinematics, with the last two having a non-vanishing on-shell limit.

We only present one helicity configuration for each process.
The opposite-helicity configuration is obtained by exchanging
%
%%%%%%%%%%%%%%%%%%%%%%%%%%%%%%%%%%%%%%%%
\begin{equation}
\kapp \leftrightarrow \kstr
\quad,\quad
\ANG{ab} \leftrightarrow \SQR{ba}
\end{equation}
%%%%%%%%%%%%%%%%%%%%%%%%%%%%%%%%%%%%%%%%
%
everywhere in the expression.
The latter also implies $\ANGSQR{a|b|c}\leftrightarrow\ANGSQR{c|b|a}$.
For the process $\procee$, the above includes the electron-positron pair.
Configurations for which only the electron and positron have flipped helicity are obtained by simply exchanging their labels $(4,5)$ everywhere in the expression.

\subsection{$\BLDempty\BLDto\BLDgSTR\BLDg\BLDg$}
In order to obtain the scattering amplitude for  $\procgg$  process we apply the auxiliary quark prescription to the  results from Appendix~II in~\cite{Bern:1994fz}.
This means to use $\procqqgg$ as the on-shell embedding process, with quark-antiquark pair taking the role of auxiliary partons.
%%%%%%%%%%%%%%%%%%%%%%%%%%%%%%%%%%%%%%%%
%
In \Appendix{App:ggauxg} we show how the method works with auxiliary gluons.

At tree-level, the color decomposition is
%
%%%%%%%%%%%%%%%%%%%%%%%%%%%%%%%%%%%%%%%%
\begin{align}
\fullAmpTree(\auxB_{\qb},\auxA_{\qq},2,3)
 = \big(T^{a_{2}}T^{a_{3}}\big)^{\imath_{\auxB}}_{i_{\auxA}}\,\AmpTree(\auxB_{\qb},\auxA_{\qq},2,3)
 + \big(T^{a_{3}}T^{a_{2}}\big)^{\imath_{\auxB}}_{i_{\auxA}}\,\AmpTree(\auxB_{\qb},\auxA_{\qq},3,2)
~.
\end{align}
%%%%%%%%%%%%%%%%%%%%%%%%%%%%%%%%%%%%%%%% 
%
Following \Equation{Eq:def04} we find
%
%%%%%%%%%%%%%%%%%%%%%%%%%%%%%%%%%%%%%%%%
%
%%%%%%%%%%%%%%%%%%%%%%%%%%%%%%%%%%%%%%%%
\begin{equation}
\fullAmpTree(1^\star,2,3)
=
 \Tr\big(T^{a_1}T^{a_2}T^{a_3}\big)\AmpTree(1^\star,2,3)
+\Tr\big(T^{a_1}T^{a_3}T^{a_2}\big)\AmpTree(1^\star,3,2)
~.
\end{equation}
%%%%%%%%%%%%%%%%%%%%%%%%%%%%%%%%%%%%%%%%
%
The helicity amplitudes are
%
%%%%%%%%%%%%%%%%%%%%%%%%%%%%%%%%%%%%%%%%
\begin{equation}
\AmpTree(1^\star,2^\pm,3^\pm) = 0
\quad,\quad
\AmpTree(1^\star,2^-,3^+) = \frac{\ANG{p2}^3}{\kstr\ANG{23}\ANG{3p}}
%\quad,\quad
%\AmpTree(1^\star,2^+,3^-) = \frac{\SQR{2p}^3}{\kapp\SQR{p3}\SQR{32}}
~.
\end{equation}
%%%%%%%%%%%%%%%%%%%%%%%%%%%%%%%%%%%%%%%%
%
Realize that $\AmpTree(1^\star,3,2)=-\AmpTree(1^\star,2,3)$ (multiply numerator and denominator by $\ANG{p2}$, and then only exchange $2\leftrightarrow3$ in the denominator).
Thus, we can also decompose the amplitude as
%
%%%%%%%%%%%%%%%%%%%%%%%%%%%%%%%%%%%%%%%%
\begin{equation}
\fullAmpTree(1^\star,2,3)
=
 \sqrt{2}\,\imag f^{a_1a_2a_3}\AmpTree(1^\star,2,3)
~.
\end{equation}
%%%%%%%%%%%%%%%%%%%%%%%%%%%%%%%%%%%%%%%%
%
The auxiliary quark one-loop color decomposition is
%
%%%%%%%%%%%%%%%%%%%%%%%%%%%%%%%%%%%%%%%%
\begin{align}
\fullAmp(\auxB_{\qb},\auxA_{\qq},2,3)
& 
= \Nc\big(T^{a_{2}}T^{a_{3}}\big)^{\imath_\auxB}_{i_\auxA}\,\Amp_{1}(\auxB_{\qb},\auxA_{\qq},2,3)
 + \Nc\big(T^{a_{3}}T^{a_{2}}\big)^{\imath_\auxB}_{i_\auxA}\,\Amp_{1}(\auxB_{\qb},\auxA_{\qq},3,2)
\notag\\&\hspace{4ex}
+\Tr\big(T^{a_{2}}T^{a_{3}}\big)\delta^{\imath_\auxB}_{i_\auxA}\,\Amp_{3}(\auxB_{\qb},\auxA_{\qq},2,3)
\notag\\&\hspace{0ex}
= \sqrt{2}\,\imag f^{a_2a_3a_1}\Nc\big(T^{a_1}\big)^{\imath_{\auxB}}_{i_{\auxA}}\Amp_{1}(\auxB_{\qb},\auxA_{\qq},2,3)
\\&\hspace{4ex}
+\Nc\big(T^{a_{3}}T^{a_{2}}\big)^{\imath_\auxB}_{i_\auxA}
     \Big[\Amp_{1}(\auxB_{\qb},\auxA_{\qq},2,3)+\Amp_{1}(\auxB_{\qb},\auxA_{\qq},3,2)\Big]
\notag\\&\hspace{4ex}
+\Tr\big(T^{a_{2}}T^{a_{3}}\big)\delta^{\imath_\auxB}_{i_\auxA}\,\Amp_{3}(\auxB_{\qb},\auxA_{\qq},2,3)
\notag~.
\end{align}
%%%%%%%%%%%%%%%%%%%%%%%%%%%%%%%%%%%%%%%% 
%
The decomposition of $\Amp_1$ into primitive amplitudes is given in Eq.~(4.2) of~\cite{Bern:1994fz}, and $\Amp_3$ in Eq.(4.6).
Regarding the latter, realize that the primitive amplitudes for which the auxiliary partons are not adjacent do not contribute.
The amplitude $\Amp_{1}(1^\star,2^+,3^+)$ does not vanish, while it corresponds to a helicity configuration that vanishes at tree-level.
It was already calculated in~\cite{Blanco:2020akb}, and can be written slightly more compactly as
%%%%%%%%%%%%%%%%%%%%%%%%%%%%%%%%%%%%%%%%
\begin{equation}
\Amp_{1}(1^\star,2^+,3^+) = \bigg(1-\frac{n_f}{\Nc}\bigg)\frac{1}{48\pi^2}\,\frac{\SQR{2p}\SQR{p3}}{\kapp\ANG{23}}
~.
\end{equation}
%%%%%%%%%%%%%%%%%%%%%%%%%%%%%%%%%%%%%%%%
%
For the other helicity amplitudes, upon taking the $\Lambda$-limit we find
%
%%%%%%%%%%%%%%%%%%%%%%%%%%%%%%%%%%%%%%%%
\begin{align}
\Amp_{3}(1^\star,2^+,3^+) &= 0
~,\\
% Using A(1^-,2^+,3^-,4^+) with p1->-Lp and p2->Lp 
\Amp_{1}(1^\star,2^-,3^+) &= \Amp^{\reLab}_{1}(1^\star,2^-,3^+)
                           - \mysign(s_{p2})\AmpRt(1^\star,2^-,3^+)
\\
\Amp_{1}(1^\star,2^-,3^+) + \Amp_{1}(1^\star,3^+,2^-) &= {-}2\mysign(s_{p2})\AmpRt(1^\star,2^-,3^+)
\\
\Amp_{3}(1^\star,2^-,3^+) &= {-}2\mysign(s_{p2})\AmpRt(1^\star,2^-,3^+)
%\cGamma\AmpTree(1^\star,2^-,3^+)\RT\bigg[-{}\frac{2}{\vep}\,\imag\pi\sgn(s_{p2})\bigg]
~,
\end{align}
%%%%%%%%%%%%%%%%%%%%%%%%%%%%%%%%%%%%%%%%
%
where $\AmpRt$ and $\mysign$ are defined in \Equation{Eq:def06} and \Equation{Eq:def07}, and
%
%%%%%%%%%%%%%%%%%%%%%%%%%%%%%%%%%%%%%%%%
\begin{align}
\Amp^{\reLab}_{1}(1^\star,2,3) &= \cGamma\AmpTree(1^\star,2,3)\RT
\notag\\&\hspace{4ex}\times
\bigg[
       \frac{-3}{\vep^2}
      +\frac{2}{\vep}\ln\frac{|s_{p2}|}{\kTsq}
      +\frac{2\pi^2}{3}-\frac{64+3\deltaR}{9}
      -\frac{11-2\nf/\Nc}{3\vep}+\frac{\nf}{\Nc}\frac{10}{9}
\bigg]
~.
\label{Eq:gggre}
\end{align}
%%%%%%%%%%%%%%%%%%%%%%%%%%%%%%%%%%%%%%%%
%
There is no discontinuity problem for $|\kT|\to0$ since $|\kT|\AmpTree$ vanishes in that limit.
%
%{\color{red}%
The helicities of the auxiliary partons were not mentioned, but either choice gives the same result.
Switching the role of $\auxA,\auxB$ results in an overall minus sign at tree-level and also for $\Amp^{\reLab}_{1}$ and $\AmpRt$, while the argument of $\mysign$ switches from $s_{p2}$ to $s_{p3}$.
Due to the kinematic restrictions, we have $s_{p3}=-s_{p2}$, and we see that the contributions proportional to $\mysign$ vanish in the combination of \Equation{symauxq}.
We can, however, already ignore those contributions because of the color content.
%}%\color{red}
%
We have
%
%%%%%%%%%%%%%%%%%%%%%%%%%%%%%%%%%%%%%%%%
\begin{align}
&\fullAmp(\auxB_{\qb},\auxA_{\qq},2,3)
\lamlim
\sqrt{2}\,\imag f^{a_{2}a_{3}a_{1}}\Nc\big(T^{a_1}\big)^{\imath_{\auxB}}_{i_{\auxA}}
\,\Amp^{\reLab}_{1}(1^\star,2,3)
\notag\\&\hspace{10ex}
-\Big[
 \Nc\big(T^{a_{2}}T^{a_{3}}\big)^{\imath_{\auxB}}_{i_{\auxA}}
+\Nc\big(T^{a_{3}}T^{a_{2}}\big)^{\imath_{\auxB}}_{i_{\auxA}}
+2\Tr\big(T^{a_{2}}T^{a_{3}}\big)\delta^{\imath_{\auxB}}_{i_{\auxA}}\Big]
  \mysign(s_{p2})\AmpRt(1^\star,2,3)
~.
\end{align}
%%%%%%%%%%%%%%%%%%%%%%%%%%%%%%%%%%%%%%%% 
%
Since the second line is symmetric in $a_2,a_3$, it vanishes when contracted with tree-level amplitude, which is anti-symmetric.
Thus, we eventually find
%
%%%%%%%%%%%%%%%%%%%%%%%%%%%%%%%%%%%%%%%%
\begin{equation}
\fullAmp^{\NLO}(1^\star,2,3) = \sqrt{2}\,\imag\,f^{a_1a_2a_3}\Nc\,\Amp^{\reLab}_{1}(1^\star,2,3)
~.
\end{equation}
%%%%%%%%%%%%%%%%%%%%%%%%%%%%%%%%%%%%%%%%
%
As mentioned before, this process vanishes in the on-shell limit $|\kT|\to0$, which is manifest through the vanishing tree-level amplitudes.

\subsection{$\BLDempty\BLDto\BLDgSTR\BLDqB\BLDq$}
%%%%%%%%%%%%%%%%%%%%%%%%%%%%%%%%%%%%%%%%
For $g^* q\bar{q}$ we can use the same on-shell embedding process as for $g^* g g$,  but considering this time the gluon pair to play the role of auxiliary partons.
%apply the auxiliary gluon method on the same amplitude as for $g^* g g$ \ie\ results from Appendix~II in~\cite{Bern:1994fz}.
%
%Let us consider auxiliary gluons.
%
At tree-level, the color decomposition is
%
%%%%%%%%%%%%%%%%%%%%%%%%%%%%%%%%%%%%%%%%
\begin{align}
\fullAmpTree(2_{\qb},3_{\qq},\auxB,\auxA)
 = \big(T^{a_{\auxB}}T^{a_{\auxA}}\big)^{\imath_2}_{i_3}\,\AmpTree(2_{\qb},3_{\qq},\auxB,\auxA)
 + \big(T^{a_{\auxA}}T^{a_{\auxB}}\big)^{\imath_2}_{i_3}\,\AmpTree(2_{\qb},3_{\qq},\auxA,\auxB)
~.
\end{align}
%%%%%%%%%%%%%%%%%%%%%%%%%%%%%%%%%%%%%%%% 
%
with
%
%%%%%%%%%%%%%%%%%%%%%%%%%%%%%%%%%%%%%%%%
\begin{align}
\AmpTree(2_{\qb},3_{\qq},\auxB,\auxA)& \lamlim \Amp^{\tree}(1^\star,2_{\qb},3_{\qq})
\\
\AmpTree(2_{\qb},3_{\qq},\auxA,\auxB)& \lamlim -\Amp^{\tree}(1^\star,2_{\qb},3_{\qq})
\end{align}
%%%%%%%%%%%%%%%%%%%%%%%%%%%%%%%%%%%%%%%%
%
so
%
%%%%%%%%%%%%%%%%%%%%%%%%%%%%%%%%%%%%%%%%
\begin{align}
&\fullAmpTree(2_{\qb},3_{\qq},\auxB,\auxA)
\lamlim
\Big[\big(T^{a_{\auxB}}T^{a_{\auxA}}\big)^{\imath_2}_{i_3}
    -\big(T^{a_{\auxA}}T^{a_{\auxB}}\big)^{\imath_2}_{i_3}\Big]\Amp^{\tree}(1^\star,2_{\qb},3_{\qq})
\notag\\&\hspace{12ex}
=\sqrt{2}\,\imag f^{a_{\auxB}a_{\auxA}a_{1}}\big(T^{a_1}\big)^{\imath_2}_{i_3}\,\Amp^{\tree}(1^\star,2_{\qb},3_{\qq})
=\sqrt{2}\,\imag f^{a_{\auxB}a_{\auxA}a_{1}}\fullAmp^{\tree,a_1}(1^\star,2_{\qb},3_{\qq})
~,
\end{align}
%%%%%%%%%%%%%%%%%%%%%%%%%%%%%%%%%%%%%%%%
%
confirming \Equation{Eq:def05}.
The helicity amplitudes are given by
\begin{equation}
\AmpTree(1^\star,2_{\qb}^+,3_{\qq}^+) = 0
\quad,\quad
\AmpTree(1^\star,2_{\qb}^-,3_{\qq}^+) = -\frac{\ANG{2p}^2}{\kstr\ANG{23}}
~.
\end{equation}
%%%%%%%%%%%%%%%%%%%%%%%%%%%%%%%%%%%%%%%%
%
The $\kstr$ in the denominator came from $\ANG{\auxB\auxA}$, and gets a minus sign when exchanging $\auxA\leftrightarrow\auxB$.
The one-loop color decomposition is
%
%%%%%%%%%%%%%%%%%%%%%%%%%%%%%%%%%%%%%%%%
\begin{align}
\fullAmp(2_{\qb},3_{\qq},\auxB,\auxA)
& = \Nc\big(T^{a_{\auxB}}T^{a_{\auxA}}\big)^{\imath_2}_{i_3}\,\Amp_{1}(2_{\qb},3_{\qq},\auxB,\auxA)
 + \Nc\big(T^{a_{\auxA}}T^{a_{\auxB}}\big)^{\imath_2}_{i_3}\,\Amp_{1}(2_{\qb},3_{\qq},\auxA,\auxB)
\notag\\&\hspace{4ex}
+\Tr\big(T^{a_{\auxA}}T^{a_{\auxB}}\big)\delta^{\imath_2}_{i_3}\,\Amp_{3}(2_{\qb},3_{\qq},\auxB,\auxA)
\\&\hspace{0ex}
= \sqrt{2}\,\imag f^{a_{\auxB}a_{\auxA}a_1}\Nc\big(T^{a_1}\big)^{\imath_2}_{i_3}\Amp_{1}(2_{\qb},3_{\qq},\auxB,\auxA)
\notag\\&\hspace{4ex}
+ \Nc\big(T^{a_{\auxA}}T^{a_{\auxB}}\big)^{\imath_2}_{i_3}
  \Big[\Amp_{1}(2_{\qb},3_{\qq},\auxB,\auxA)+\Amp_{1}(2_{\qb},3_{\qq},\auxA,\auxB)\Big]
\notag\\&\hspace{4ex}
+\Tr\big(T^{a_{\auxA}}T^{a_{\auxB}}\big)\delta^{\imath_2}_{i_3}\,\Amp_{3}(2_{\qb},3_{\qq},\auxB,\auxA)
~.
\end{align}
%%%%%%%%%%%%%%%%%%%%%%%%%%%%%%%%%%%%%%%% 
%
The expression of $\Amp_1$ into primitive amplitudes is given in Eq.~(4.2) of~\cite{Bern:1994fz}, and $\Amp_3$ in Eq.(4.6).
Regarding the latter, realize that the primitive amplitudes for which the auxiliary partons are not adjacent do not contribute.
We must choose opposite helicities for the auxiliary gluons, and may choose $\auxB^-,\auxA^+$.
For the helicity amplitudes after the $\Lambda$-limit we find
\begin{align}
\Amp_{1}(1^\star,2_{\qb}^+,3_{\qq}^+) &= \Amp_{3}(1^\star,2_{\qb}^+,3_{\qq}^+) = 0
~,\\
\Amp_{1}(1^\star,2_{\qb}^-,3_{\qq}^+) &= \Amp^{\reLab}_{1}(1^\star,2_{\qb}^-,3_{\qq}^+) 
                           - \mysign(s_{p2})\AmpRt(1^\star,2_{\qb}^-,3_{\qq}^+) ~,
\\
\Amp_{3}(1^\star,2_{\qb}^-,3_{\qq}^+) &= -2\mysign(s_{p2})\AmpRt(1^\star,2_{\qb}^-,3_{\qq}^+)
\\
\Amp_{1}(2_{\qb}^-,3_{\qq}^+,\auxB^-,\auxA^+)+\Amp_{1}(2_{\qb}^-,3_{\qq}^+,\auxA^+,\auxB^-) &\lamlim -2\mysign(s_{p2})\AmpRt(1^\star,2_{\qb}^-,3_{\qq}^+) 
~,
\end{align}
%%%%%%%%%%%%%%%%%%%%%%%%%%%%%%%%%%%%%%%%
%
where $\AmpRt$ and $\mysign$ are defined in \Equation{Eq:def06} and \Equation{Eq:def07}, and
\begin{align}
&\Amp^{\reLab}_{1}(1^\star,2_{\qb},3_{\qq})
 = \cGamma\AmpTree(1^\star,2_{\qb},3_{\qq})\RT
\notag\\&\hspace{8ex}\times
\bigg[
  \frac{-2}{\vep^2}
  +\frac{1}{\vep}\bigg(2\ln\frac{|s_{p2}|}{\kTsq}-\frac{3}{2}\bigg)
  +\frac{2\pi^2}{3}-\frac{5+\deltaR}{2}
  +\frac{1}{\Nc^2}\bigg(\frac{1}{\vep^2}+\frac{3}{2\vep}+\frac{7+\deltaR}{2}\bigg)
\bigg]
\label{Eq:gqqre}
~.
\end{align}
%
%{\color{red}%
With the same reasoning as for $\procqq$, we see that the contributions proportional to $\mysign$ vanish in the combination of \Equation{symauxg}.
Also again, those contributions are already irrelevant due to the color content.
%}%\color{red}
Combining, we find
%
%%%%%%%%%%%%%%%%%%%%%%%%%%%%%%%%%%%%%%%%
\begin{align}
&\fullAmp(2_{\qb},3_{\qq},\auxB,\auxA)
\lamlim
\sqrt{2}\,\imag f^{a_{\auxB}a_{\auxA}a_1}\Nc\big(T^{a_1}\big)^{\imath_2}_{i_3}
\,\Amp^{\reLab}_{1}(1^\star,2_{\qb},3_{\qq})
\notag\\&\hspace{8ex}
-\bigg[
 \Nc\big(T^{a_{\auxA}}T^{a_{\auxB}}\big)^{\imath_2}_{i_3}
+\Nc\big(T^{a_{\auxB}}T^{a_{\auxA}}\big)^{\imath_2}_{i_3}
+2\Tr\big(T^{a_{\auxA}}T^{a_{\auxB}}\big)\delta^{\imath_2}_{i_3}\bigg]
  \mysign(s_{p2})\AmpRt(1^\star,2_{\qb},3_{\qq})
~.
\end{align}
%%%%%%%%%%%%%%%%%%%%%%%%%%%%%%%%%%%%%%%% 
%
Since the second line is symmetric in $a_2,a_3$, it vanishes when contracted with tree-level amplitude, which is anti-symmetric.
Thus, we find
%
%%%%%%%%%%%%%%%%%%%%%%%%%%%%%%%%%%%%%%%%
\begin{equation}
\fullAmp^{\NLO}(1^\star,2_{\qb},3_{\qq})
= \Nc\big(T^{a_1}\big)^{\imath_2}_{i_3}
\,\Amp^{\reLab}_{1}(1^\star,2_{\qb},3_{\qq})
~.
\end{equation}
%%%%%%%%%%%%%%%%%%%%%%%%%%%%%%%%%%%%%%%%
%
Also this process vanishes in the on-shell limit $|\kT|\to0$, which is manifest through the vanishing tree-level amplitudes.

\subsection{$\BLDempty\BLDto\BLDgSTR\BLDg\BLDh$}
For this process the color decomposition is trivial and there is only one partial amplitude.
Following \cite{Schmidt:1997wr}, with auxiliary quarks we have both at tree-level and one-loop
%
%%%%%%%%%%%%%%%%%%%%%%%%%%%%%%%%%%%%%%%%
\begin{align}
\fullAmp^{\tree}(\auxA_{\qq},\auxB_{\qb},2,H) &\lamlim
 -\frac{\alphaS}{3\pi v}\frac{1}{2}\big(T^{a_2}\big)^{\imath_{\auxB}}_{i_{\auxA}}
\,\Amp^{\tree}(1^\star,2,H)
~,\\
\fullAmp(\auxA_{\qq},\auxB_{\qb},2,H) &\lamlim
 -\frac{\alphaS}{3\pi v}\frac{1}{2}\,\Nc\big(T^{a_2}\big)^{\imath_{\auxB}}_{i_{\auxA}}
\,\Amp(1^\star,2,H)
~,
\end{align}
%%%%%%%%%%%%%%%%%%%%%%%%%%%%%%%%%%%%%%%%
%
that is
%
%%%%%%%%%%%%%%%%%%%%%%%%%%%%%%%%%%%%%%%%
\begin{align}
\fullAmp^{\tree}(1^\star,2,H)
&=
-\frac{\alphaS}{3\pi v}\frac{1}{2}\,\delta^{a_1a_2}
\,\Amp^{\tree}(1^\star,2,H)
~,\\
\fullAmp(1^\star,2,H) &=
 -\frac{\alphaS}{3\pi v}\frac{1}{2}\,\Nc\delta^{a_1a_2}
\,\Amp(1^\star,2,H)
~.
\end{align}
%%%%%%%%%%%%%%%%%%%%%%%%%%%%%%%%%%%%%%%%
%
The factor $-\alphaS/(3\pi v)$ is related to the effective Higgs-gluon coupling.
At tree-level, we get
%
%%%%%%%%%%%%%%%%%%%%%%%%%%%%%%%%%%%%%%%%
\begin{equation}
\AmpTree(1^\star,2^+,H) = \frac{\SQR{p2}^2}{\kapp}
\end{equation}
%%%%%%%%%%%%%%%%%%%%%%%%%%%%%%%%%%%%%%%%
%
and at one loop we get
%
%%%%%%%%%%%%%%%%%%%%%%%%%%%%%%%%%%%%%%%%
\begin{align}
\Amp(1^\star,2,H) &= 
\cGamma\AmpTree(1^\star,2,H)\bigg[
   {-}\frac{2}{\vep^2}\left(\frac{\mu^2}{-M_H^2}\right)^\vep
   + \frac{2}{\vep}\RT\ln\frac{|s_{p2}|}{-M_H^2}
\notag\\&\hspace{38ex}
   + \frac{\pi^2}{3}-2\mathrm{Li}_2\left(1+\frac{\kTsq}{M_H^2}\right)
\bigg]
~.
\end{align}
%%%%%%%%%%%%%%%%%%%%%%%%%%%%%%%%%%%%%%%%
%
Here we must imagine that $M_H^2$ has a small positive imaginary part (so it plays the role of an invariant rather than a squared mass, which would naturally get a negative imaginary part).
Notice that the first term in the square brackets gives the on-shell result, while the other terms vanish for $|\kT|\to0$ (and negative $M_H^2$ implying $-M_H^2\to|s_{p2}|$).
The following organization of terms better matches the singularity structure of the real radiation contribution in an NLO calculation:
%
%%%%%%%%%%%%%%%%%%%%%%%%%%%%%%%%%%%%%%%%
\begin{align}
%\Amp(1^\star,2^+,H) &= 
%\cGamma\AmpTree(1^\star,2^+,H)\bigg[
%   {-}\frac{2}{\vep^2}\left(\frac{\mu^2}{|s_{p2}|}\right)^\vep
%   + \ln\frac{|s_{p2}|}{-M_H^2}\bigg(
%      \ln\frac{|s_{p2}|}{\kTsq}+\ln\frac{-M_H^2}{\kTsq}\bigg)
%\notag\\&\hspace{40ex}
%   + \frac{\pi^2}{3}-2\mathrm{Li}_2\left(1+\frac{\kTsq}{M_H^2}\right)
%\bigg]
%~.
\Amp(1^\star,2,H) &= 
\cGamma\AmpTree(1^\star,2,H)\bigg\{
   {-}\frac{2}{\vep^2}\left(\frac{\mu^2}{|s_{p2}|}\right)^\vep
   + \ln\bigg(\frac{|s_{p2}|}{|M_H^2|}\bigg)
     \ln\bigg(\frac{|s_{p2}||M_H^2|}{\kTsq^2}\bigg)
\notag\\&\hspace{10ex}
   + \frac{\pi^2}{3}-2\mathrm{Li}_2\left(1+\frac{\kTsq}{M_H^2}\right)
   + \theta\big(M_H^2\big)\bigg[\pi^2 + 2\imag\pi\ln\bigg(\frac{|M_H^2|}{\kTsq}\bigg)\bigg]
\bigg\}
~.
\end{align}
%%%%%%%%%%%%%%%%%%%%%%%%%%%%%%%%%%%%%%%%
%

%%%%%%%%%%%%%%%%%%%%%%%%%%%%%%%%%%%%%%%%%%%%%%%%%%%%%%%%%%%%%%%%%%%%%%%%%%%%%%%%
%%% 0 -> g^* qBar q e+ e- %%%%%%%%%%%%%%%%%%%%%%%%%%%%%%%%%%%%%%%%%%%%%%%%%%%%%%
%%%%%%%%%%%%%%%%%%%%%%%%%%%%%%%%%%%%%%%%%%%%%%%%%%%%%%%%%%%%%%%%%%%%%%%%%%%%%%%%
\subsection{$\BLDempty\BLDto\BLDgSTR\BLDqB\BLDq\BLDePLS\BLDeMIN$}
In this case, we use the results from \cite{Bern:1997sc}, and consider auxiliary quarks.
We are after the amplitude in  Eq.~(2.11) of that publication, that is
%
%%%%%%%%%%%%%%%%%%%%%%%%%%%%%%%%%%%%%%%%
\begin{align}
&\fullAmp^{\mathrm{1-loop}}(1_{\qq},2_{\QB},3_{\QQ},4_{\qb},4_{\ePLS},5_{\eMIN})
\notag\\&\hspace{0ex}
= 2\gQED^2\big({-}Q^q+v^e_{L,R}v^q_{L,R}\EuScript{P}_Z(s_{45})\big)
  \bigg[
    \Nc\delta^{\imath_{\auxA}}_{i_{3}}\delta^{\imath_{2}}_{i_{\auxB}}
    \Amp_{1}(1_{\qq},2_{\QB},3_{\QQ},4_{\qb})
    +\delta^{\imath_{2}}_{i_{3}}\delta^{\imath_{\auxA}}_{i_{\auxB}}
    \Amp_{2}(1_{\qq},2_{\QB},3_{\QQ},4_{\qb})
  \bigg]
\notag\\&\hspace{0ex}
+ 2\gQED^2\big({-}Q^Q+v^e_{L,R}v^Q_{L,R}\EuScript{P}_Z(s_{45})\big)
  \bigg[
    \Nc\delta^{\imath_{\auxA}}_{i_{3}}\delta^{\imath_{2}}_{i_{\auxB}}
    \Amp_{1}(3_{\QQ},4_{\qb},1_{\qq},2_{\QB})
    +\delta^{\imath_{2}}_{i_{3}}\delta^{\imath_{\auxA}}_{i_{\auxB}}
    \Amp_{2}(3_{\QQ},4_{\qb},1_{\qq},2_{\QB})
  \bigg]
\notag\\&\hspace{18ex}
+ 2\gQED^2\frac{v^e_{L,R}}{\sin2\theta_W}\,\EuScript{P}_Z(s_{45})
  \bigg(\delta^{\imath_{\auxA}}_{i_{3}}\delta^{\imath_{2}}_{i_{\auxB}}
        -\frac{1}{\Nc}\delta^{\imath_{2}}_{i_{3}}\delta^{\imath_{\auxA}}_{i_{\auxB}}\bigg)
   \Amp_{3}(1_{\qq},2_{\QB},3_{\QQ},4_{\qb})
~.
\end{align}
On the left-hand side, the enumeration of the $\ePLS,\eMIN$ pair is written explicitly, while it is omitted in the partial amplitudes.
We will write it explicitly in the partial amplitudes obtained after the $\Lambda$ limit.
One issue to keep in mind is that in our prescription, {\em the vector-boson must not couple to the auxiliary quark line}.
This is achieved by including only the contributions of Figure~3 in~\cite{Bern:1997sc}, while assigning $Q\bar{Q}$ as the auxiliary quark-antiquark pair.
This simply means that we must exclude the amplitudes with ordering $\big(3_{Q},4_{\bar{q}},1_{q},2_{\bar{Q}}\big)$ in the expression above.
Thus, we need
%
%%%%%%%%%%%%%%%%%%%%%%%%%%%%%%%%%%%%%%%%
\begin{align}
\fullAmp(3_{\qq},\auxA_{\QB},\auxB_{\QQ},2_{\qb},4_{\ePLS},5_{\eMIN})
&= \Cone
  \bigg[
    \Nc\delta^{\imath_{\auxA}}_{i_{3}}\delta^{\imath_{2}}_{i_{\auxB}}
    \Amp_{1}(3_{\qq},\auxA_{\QB},\auxB_{\QQ},2_{\qb})
    +\delta^{\imath_{2}}_{i_{3}}\delta^{\imath_{\auxA}}_{i_{\auxB}}
    \Amp_{2}(3_{\qq},\auxA_{\QB},\auxB_{\QQ},2_{\qb})
  \bigg]
\notag\\&\hspace{0ex}
+ \Cthree
  \bigg(\delta^{\imath_{\auxA}}_{i_{3}}\delta^{\imath_{2}}_{i_{\auxB}}
        -\frac{1}{\Nc}\delta^{\imath_{2}}_{i_{3}}\delta^{\imath_{\auxA}}_{i_{\auxB}}\bigg)
   \Amp_{3}(3_{\qq},\auxA_{\QB},\auxB_{\QQ},2_{\qb})
~,
\end{align}
%%%%%%%%%%%%%%%%%%%%%%%%%%%%%%%%%%%%%%%%
%
where we abbreviate
%
%%%%%%%%%%%%%%%%%%%%%%%%%%%%%%%%%%%%%%%%
\begin{equation}
\Cone = 2\gQED^2\big({-}Q^q+v^e_{L,R}v^q_{L,R}\EuScript{P}_Z(s_{45})\big)
\quad,\quad
\Cthree = 2\gQED^2\frac{v^e_{L,R}}{\sin2\theta_W}\,\EuScript{P}_Z(s_{45})
~.
\end{equation}
%%%%%%%%%%%%%%%%%%%%%%%%%%%%%%%%%%%%%%%%
%
We copied the symbols for the constants, and refer to page~7 of~\cite{Bern:1997sc} for their definition.
We only mention that the partial amplitudes include the photon propagator, and the factor $\EuScript{P}_Z(s_{45})$ corrects it to the $Z$-boson propagator, so both contributions are included depending on whether the constants $v^{e,q}_{L,R}$ are switched on.
%
%On the right-hand side, $\gQED$ is the QED coupling, $\gQCD$ the QCD coupling, $Q^q$ is the charge of quark $\qq$, and
%%%%%%%%%%%%%%%%%%%%%%%%%%%%%%%%%%%%%%%%%
%\begin{equation}
%\EuScript{P}_Z(s) = \frac{s}{s-M_Z^2+\imag\Gamma_ZM_Z}
%\end{equation}
%%%%%%%%%%%%%%%%%%%%%%%%%%%%%%%%%%%%%%%%%
%is the ratio of Z and photon propagators, with $M_Z$ and $\Gamma_Z$ are the mass and width of the Z-boson.
%
The tree-level decomposition is given by
%
%
%%%%%%%%%%%%%%%%%%%%%%%%%%%%%%%%%%%%%%%%
\begin{equation}
\fullAmpTree(3_{\qq},\auxA_{\QB},\auxB_{\QQ},2_{\qb},4_{\ePLS},5_{\eMIN})
= \Cone
  \bigg(\delta^{\imath_{\auxA}}_{i_{3}}\delta^{\imath_{2}}_{i_{\auxB}}
        -\frac{1}{\Nc}\delta^{\imath_{2}}_{i_{3}}\delta^{\imath_{\auxA}}_{i_{\auxB}}\bigg)
    \AmpTree(3_{\qq},\auxA_{\QB},\auxB_{\QQ},2_{\qb})
~.
\end{equation}
%%%%%%%%%%%%%%%%%%%%%%%%%%%%%%%%%%%%%%%%
%
Notice that the tree-level expression vanishes under contraction with $\delta_{\imath_{\auxA}}^{i_{\auxB}}$.
We can write the one-loop amplitude in terms of the tree-level color structure, and one that vanishes if contracted with the tree-level structure as
\begin{align}
\fullAmp(3_{\qq},\auxA_{\QB},\auxB_{\QQ},2_{\qb},4_{\ePLS},5_{\eMIN})
&=
  \bigg(\delta^{\imath_{\auxA}}_{i_{3}}\delta^{\imath_{2}}_{i_{\auxB}}
        -\frac{1}{\Nc}\delta^{\imath_{2}}_{i_{3}}\delta^{\imath_{\auxA}}_{i_{\auxB}}\bigg)\bigg[
  \Cone\Nc\Amp_{1}(3_{\qq},\auxA_{\QB},\auxB_{\QQ},2_{\qb})
\notag\\&\hspace{26ex}
+ \Cthree\Amp_{3}(3_{\qq},\auxA_{\QB},\auxB_{\QQ},2_{\qb})
  \bigg]
\notag\\&\hspace{0ex}
+\Cone\,\delta^{\imath_{2}}_{i_{3}}\delta^{\imath_{\auxA}}_{i_{\auxB}}
\bigg[\Amp_{1}(3_{\qq},\auxA_{\QB},\auxB_{\QQ},2_{\qb})+\Amp_{2}(3_{\qq},\auxA_{\QB},\auxB_{\QQ},2_{\qb})\bigg]
\label{Eq:qqee04}
~.
\end{align}
The contraction of color indices of the tree-level amplitude with the last line vanishes.
Omitting this term and contracting with $(T^{a_1})_{\imath_{\auxA}}^{i_{\auxB}}$, we find
%
%
%%%%%%%%%%%%%%%%%%%%%%%%%%%%%%%%%%%%%%%%
\begin{align}
\fullAmp^{\tree}(1^\star,2_{\qb},3_{\qq},4_{\ePLS},5_{\eMIN})
&= (T^{a_1})^{i_2}_{i_3}
  \,\Cone\,\Amp^{\tree}(1^\star,2_{\qb},3_{\qq},4_{\ePLS},5_{\eMIN})
~,
\\
\fullAmp^{\NLO}(1^\star,2_{\qb},3_{\qq},4_{\ePLS},5_{\eMIN})
&=
  (T^{a_1})^{i_2}_{i_3}
  \,\Big[
  \Cone\Nc\Amp_{1}(1^\star,2_{\qb},3_{\qq},4_{\ePLS},5_{\eMIN})
\notag\\&\hspace{12ex}
+ \Cthree\Amp_{3}(1^\star,2_{\qb},3_{\qq},4_{\ePLS},5_{\eMIN})
  \Big]
~.
\label{Eq:qqee01}
\end{align}
%%%%%%%%%%%%%%%%%%%%%%%%%%%%%%%%%%%%%%%%
%
We only present expressions for the helicity amplitude $\fullAmp(1^\star,2_{\qb}^-,3_{\qq}^+,4_{\ePLS}^-,5_{\eMIN}^+)$.
The others can be derived with the rules mentioned at the beginning of this section.

The partial amplitudes are decomposed into primitive amplitudes in~\cite{Bern:1997sc} as
%
%%%%%%%%%%%%%%%%%%%%%%%%%%%%%%%%%%%%%%%%
\begin{align}
\Amp_{1}(3_{\qq}^+,\auxA_{\QB}^+,\auxB_{\QQ}^-,2_{\qb}^-) &= 
   \Amp^{++}(3,\auxA,\auxB,2)
  + \frac{1}{\Nc^2}\,\Amp^{\mathrm{sl}}(\auxA,\auxB,3,2)
\notag\\&\hspace{0ex}
  - \frac{2}{\Nc^2}\big[\Amp^{++}(3,\auxA,\auxB,2)+\Amp^{+-}(3,\auxB,\auxA,2)\big]
\label{Eq:qqee05}
\\&\hspace{0ex}
  - \frac{n_f}{\Nc}\big[\Amp^{f,++}(3,\auxA,\auxB,2)+\Amp^{s,++}(3,\auxA,\auxB,2)\big]
  + \frac{1}{\Nc}\,\Amp^{t,++}(3,\auxA,\auxB,2)
%  + \frac{2\cGamma}{15}\frac{\kTsq}{m_t^2}\,\Amp^{\tree,++}(3,\auxA,\auxB,2)
\notag~,
\end{align}
%%%%%%%%%%%%%%%%%%%%%%%%%%%%%%%%%%%%%%%%
and
%
%%%%%%%%%%%%%%%%%%%%%%%%%%%%%%%%%%%%%%%%
\begin{align}
\Amp_{1}(3_{\qq}^+,\auxA_{\QB}^-,\auxB_{\QQ}^+,2_{\qb}^-) &= 
   \Amp^{+-}(3,\auxA,\auxB,2)
  - \frac{1}{\Nc^2}\,\Amp^{\mathrm{sl}}(\auxB,\auxA,3,2)
\notag\\&\hspace{0ex}
  - \frac{2}{\Nc^2}\big[\Amp^{+-}(3,\auxA,\auxB,2)+\Amp^{++}(3,\auxB,\auxA,2)\big]
\label{Eq:qqee06}
\\&\hspace{0ex}
  - \frac{n_f}{\Nc}\big[\Amp^{f,+-}(3,\auxA,\auxB,2)+\Amp^{s,+-}(3,\auxA,\auxB,2)\big]
  + \frac{1}{\Nc}\,\Amp^{t,+-}(3,\auxA,\auxB,2)
\notag~.
\end{align}
%%%%%%%%%%%%%%%%%%%%%%%%%%%%%%%%%%%%%%%%
%
From now on, we omit the top quark vacuum polarization contribution $\Amp^{t}$, since it is suppressed by a factor $\kTsq/m_t^2$ already at the amplitude level.
Also, we will not mention the $n_f/\Nc$ contribution from massless fermion loops any further, because it turns out to be completely given by the one in \Equation{oneloopauxq} and does not show up in $\Amp_1(1^\star,2_{\qb},3_{\qq},4_{\ePLS},5_{\eMIN})$.
%

%{\color{red}%
The second line \Equation{Eq:qqee05} turns out to lead to a contribution that causes a difference compared to auxiliary gluons.
In the sum over helicities of the auxiliary partons following \Equation{Eq:virthelsum}, that is by and adding \Equation{Eq:qqee06}, this $1/\Nc^2$ contribution becomes symmetric in $\auxA,\auxB$.
Including then the contribution with the role of $\auxA,\auxB$ switched as in \Equation{symauxq}, this contribution vanishes.

The primitive amplitudes in~\cite{Bern:1997sc} are decomposed further as (remember that we took out an overall factor $\imag$ compared to ~\cite{Bern:1997sc})
%
%%%%%%%%%%%%%%%%%%%%%%%%%%%%%%%%%%%%%%%%
\begin{equation}
\Amp = \cGamma\big( \Vamp\,\Amp^{\tree} + \Famp \big)
~,
\end{equation}
%%%%%%%%%%%%%%%%%%%%%%%%%%%%%%%%%%%%%%%%
%
where $\Vamp$ contains all $1/\vep$ poles and $\Famp$ is finite.
This decomposition is not uniquely defined since $\Vamp$ also contains finite parts.
We will also present the results in this type of decomposition, but with a distribution of finite terms between $\Vamp\,\Amp^{\tree}$ and $\Famp$ that does not directly correspond to the one in~\cite{Bern:1997sc}.
The cancellation of terms proportional to $\ln\Lambda$ and $\ln^2\Lambda$ happens between $\Vamp\,\Amp^{\tree}$ and $\Famp$ from~\cite{Bern:1997sc} and does not stay within each of these contributions.
The subtraction dictated by \Equation{oneloopauxq}, however, does apply only to the contribution $\Vamp\,\Amp^{\tree}$.
We write
%
%%%%%%%%%%%%%%%%%%%%%%%%%%%%%%%%%%%%%%%%
\begin{align}
\cGamma^{-1}\Amp^{++}(3,\auxA,\auxB,2) &\lamlim
   \Vamp^{\,\star}\Amp^{\tree,\star}
  +\Famp^{\,\mathrm{re},\star} 
  + \mysign(s_{p2})\Famp^{\,\mathrm{im},\star}
  +\Theta(s_{p2})\Amp^{\mathrm{Rt},\star}  
~,\notag\\
\cGamma^{-1}\Amp^{\mathrm{sl}}(\auxA,\auxB,3,2) &\lamlim
   {-}\Vamp^{\,\mathrm{sl},\star} \Amp^{\tree,\star} +\Famp^{\,\mathrm{sl},\star}
\label{Eq:qqee03}
~,
\end{align}
%%%%%%%%%%%%%%%%%%%%%%%%%%%%%%%%%%%%%%%%
%
where we use the single superscript $\star$ instead of the argument list $(1^\star,2_{\qb}^-,3_{\qq}^+,4_{\ePLS}^-,5_{\eMIN}^+)$ for brevety, and $\Theta,\mysign$ are defined in \Equation{Eq:def07}.
In the following, we present the expressions for the amplitudes on the right hand sides.
At tree level, we have
%
%%%%%%%%%%%%%%%%%%%%%%%%%%%%%%%%%%%%%%%%
\begin{align}
\Amp^{\tree,\star}
%&=
%\frac{-1}{\kapp\kstr s_{45}}\bigg[
%\frac{\SQR{3p}\ANG{42}\big(\ANG{p3}\SQR{35}+\kstr\SQR{p5}\big)}{s_{k3}}
%+
%\frac{\ANG{p2}\SQR{53}\big(\kapp \ANG{4p}+\ANG{42}\SQR{2p}\big)}{s_{k2}}
%\bigg]
%\notag\\&\hspace{2ex}
&=
\frac{-1}{s_{45}}\bigg[
  \frac{\SQR{3p}\ANG{42}\SQR{p5}}{\kapp s_{k3}}
  +
  \frac{\ANG{p2}\SQR{53}\ANG{4p}}{\kstr s_{k2}}
  +
  \frac{\ANG{24}\SQR{53}}{\kapp\kstr}\bigg(
    \frac{s_{p3}}{s_{k3}}
   -\frac{s_{p2}}{s_{k2}}
  \bigg)
\bigg]
~.
\end{align}
%%%%%%%%%%%%%%%%%%%%%%%%%%%%%%%%%%%%%%%%
%
Realize that
%
%%%%%%%%%%%%%%%%%%%%%%%%%%%%%%%%%%%%%%%%
\begin{equation}
s_{ki} \;\overset{|\kT|\to0}{\longrightarrow}\; s_{pi}
\qquad\textrm{so}\qquad
    \frac{s_{p3}}{s_{k3}}
   -\frac{s_{p2}}{s_{k2}}
\;\overset{|\kT|\to0}{\longrightarrow}\;0
\end{equation}
%%%%%%%%%%%%%%%%%%%%%%%%%%%%%%%%%%%%%%%%
%
and the amplitude behaves only as $|\kT|^{-1}$, not as $|\kT|^{-2}$.
Another form in which this is explicit is given by
\begin{align}
\Amp^{\tree,\star}
= -\frac{\Tamp^{\,\flat} + \Tamp^{\,\sharp} }{2}
~,
\end{align}
where
%
%%%%%%%%%%%%%%%%%%%%%%%%%%%%%%%%%%%%%%%%
\begin{equation}
\Tamp^{\,\flat}
=
  \frac{\SQR{3p}^2\ANG{24}^2}{\kapp \ANG{45}s_{k3}\ANGSQR{2|k|3}}
 +\frac{\ANGSQR{p|(2+4)|5}^2}{\kstr\SQR{54}s_{k3}\ANGSQR{3|k|2}}
~,
\end{equation}
%%%%%%%%%%%%%%%%%%%%%%%%%%%%%%%%%%%%%%%%%%%%%%%%%%%%%%%%%%%%%%%%%%%%%%%%%%%%%%%
%
and where $\Tamp^{\,\sharp}$ is obtained from $\Tamp^{\,\flat}$ with the ``flip'' operation, defined through the rules
%
%%%%%%%%%%%%%%%%%%%%%%%%%%%%%%%%%%%%%%%%%
\begin{align}
&\mathrm{flip}:
\;\; 2\leftrightarrow3
\;\;,\quad 4\leftrightarrow5
\;\;,\quad \ANG{ab}\leftrightarrow\SQR{ab}
\;\;,\quad \kapp\leftrightarrow\kstr
\;\;,\quad \rho_{\mathrm{flip}}=1\rightarrow-1
\;\; .
\end{align}
%%%%%%%%%%%%%%%%%%%%%%%%%%%%%%%%%%%%%%%%
%
The third rule also implies $\ANGSQR{a|b|c}\leftrightarrow\SQRANG{a|b|c}=\ANGSQR{c|b|a}$, and the last rule will show to be usefull at one loop.
Notice that each ${-}\Tamp^{\,\flat}$ and ${-}\Tamp^{\,\sharp}$ produce the on-shell limit when $|\kT|\to0$, with the $1/\kstr$ term giving the minus helicity, and the $1/\kapp$ term giving the plus helicity.
For the one-loop amplitude, we find
%
%%%%%%%%%%%%%%%%%%%%%%%%%%%%%%%%%%%%%%%%
\begin{equation}
\Vamp^{\,\star} = -\frac{1}{\vep^2}\bigg[
    \bigg(\frac{\mu^2}{-s_{p3}}\bigg)^\vep
   +\bigg(\frac{\mu^2}{-s_{p2}}\bigg)^\vep
  \bigg]
  - \frac{3}{2\vep}\bigg(\frac{\mu^2}{-s_{45}}\bigg)^\vep
   - \frac{7}{2}-\frac{\delta_R}{3}
~.
\end{equation}
%%%%%%%%%%%%%%%%%%%%%%%%%%%%%%%%%%%%%%%%
%
We included the leading-color contribution of
%
%%%%%%%%%%%%%%%%%%%%%%%%%%%%%%%%%%%%%%%%
\begin{equation}
-\frac{\delta_R}{2}\bigg(1-\frac{1}{\Nc^2}\bigg)
\label{Eq:deltaR}
\end{equation}
%%%%%%%%%%%%%%%%%%%%%%%%%%%%%%%%%%%%%%%%
%
to switch the amplitude between the 't~Hooft-Velman scheme ($\delta_R=1$) and the four-dimensional helicity scheme ($\delta_R=0)$, and subtracted the necessary leading color terms following \Equation{oneloopauxq}.
For the remaining finite parts of the amplitude we find
\begin{align}
\Famp^{\,\mathrm{re},\star} &=
\Tamp^{\,\flat}
\bigg[\ln\bigg(\frac{\kTsq}{-s_{k3}}\bigg)\ln\bigg(\frac{-s_{p3}}{-s_{k3}}\bigg)
       -\frac{1}{2}\ln^2\bigg(\frac{-s_{p2}}{-s_{k3}}\bigg)
\notag\\&\hspace{8ex}
       +\frac{1}{2}\ln\bigg(\frac{-s_{p2}}{-s_{45}}\bigg)
          \ln\bigg(\frac{-s_{p2}}{-s_{p3}}\bigg)
       + \mathrm{Li}_2\bigg(1 + \frac{\kTsq}{s_{k3}}\bigg)
       - \frac{\pi^2}{4}
       - \mathrm{Li}_2\bigg(1 - \frac{s_{45}}{s_{k3}}\bigg)
\bigg]
\notag\\&
+ 2 \frac{ \ANGSQR{p|(2+4)|5} }{s_{k3}\SQR{45} \ANGSQR{3|k|2} } 
    \bigg[\rho_{\mathrm{flip}}\,\frac{ \ANGSQR{3|k|5} \SQR{3p} }{ s_{k3} }\,\mathrm{L}_0\bigg(\frac{ \kTsq }{ -s_{k3} }\bigg) 
     - \frac{ \ANGSQR{p|2|5} }{ \kstr }\,\mathrm{L}_0\bigg(\frac{ -s_{45} }{ -s_{k3} }\bigg)
    \bigg]
\\&
-\frac{ 1 }{ 2 } \frac{\ANGSQR{3|k|5}^2 \SQR{3p}^2}{ \kapp \SQR{45} s_{k3}^2 \ANGSQR{3|k|2} }
  \,\bigg[\frac{\kTsq}{s_{k3}}\,\mathrm{L}_1\bigg(\frac{ \kTsq}{ -s_{k3} }\bigg)-1\bigg]
   + \frac{ 1 }{ 2 } \frac{\ANGSQR{p|2|5}^2}{ \kstr \SQR{45} s_{k3} \ANGSQR{3|k|2} }\,\mathrm{L}_1\bigg(\frac{ -s_{45} }{ -s_{k3} } \bigg)
\notag\\&\hspace{0ex}
+\phantom{\bigg[}\mathrm{flip}
\notag~,
\end{align}
where $\mathrm{flip}$ acts on all preceding terms.
For the other one, we find
%
%
%%%%%%%%%%%%%%%%%%%%%%%%%%%%%%%%%%%%%%%%
\begin{equation}
\Famp^{\,\mathrm{im},\star} =
  \frac{\Tamp^{\,\flat}-\Tamp^{\,\sharp}}{2}\ln\bigg(\frac{(-s_{p2})(-s_{p3})}{\kTsq(-s_{45})}\bigg)
  +\Tamp^{\,\flat}\ln\bigg(\frac{-s_{k3}}{-s_{p3}}\bigg)
  -\Tamp^{\,\sharp}\ln\bigg(\frac{-s_{k2}}{-s_{p2}}\bigg)
~.
\end{equation}
%%%%%%%%%%%%%%%%%%%%%%%%%%%%%%%%%%%%%%%%
%
While the quantities $\Famp^{\,\mathrm{re},\star}$ and $\mysign(s_{p2})\Famp^{\,\mathrm{im},\star}$ are not strictly real and imaginary, we decide to still use these labels to distinguish them from each other.
Notice that the arguments of the logarithms in $\Famp^{\,\mathrm{im},\star}$ never become negative: first of all $s_{ki}$ and $s_{pi}$ always have the same sign. For $e^+,e^-$ in the final state, one of $p_{2,3}$ must be in the initial state while the other is in the final state and we have $s_{45}>0$ and $s_{p2}s_{p3}<0$.
For DIS-type kinematics we have $s_{45}<0$, $s_{p2}<0$, $s_{p3}<0$.

Also notice that $\Famp^{\,\mathrm{im},\star}$ vanishes in the on-shell limit.
The terms proportional to $1/\kapp$ in $\Vamp^{\,\star}\Amp^{\mathrm{tree},\star}+\Famp^{\,\mathrm{re},\star}$ exactly give the on-shell limit for $|\kT|\to0$, as given in the formulas in Appendix~IV in~\cite{Bern:1997sc}, which involve a positive-helicity gluon.
For the combination of amplitudes in \Equation{symauxq} we find
%
%%%%%%%%%%%%%%%%%%%%%%%%%%%%%%%%%%%%%%%%
\begin{equation}
4\big(\Vamp^{\,\star}\Amp^{\mathrm{tree},\star}+\Famp^{\,\mathrm{re},\star}\big)
+2\big[\mysign(s_{p2})-\mysign(s_{p3})\big]\Famp^{\,\mathrm{im},\star}
+2\big[\Theta(s_{p2})+\Theta(s_{p3})\big]\Amp^{\mathrm{Rt},\star}
~.
\end{equation}
%%%%%%%%%%%%%%%%%%%%%%%%%%%%%%%%%%%%%%%%
%
Notice that the ``imaginary'' terms vanish for DIS-type kinematics.
The term with $\Amp^{\mathrm{Rt},\star}$ is truly imaginary, and never contributes to the virtual contribution.
%}%\color{red}

The $1/\Nc^2$ contribution does not contain such pieces, and we find
%
%%%%%%%%%%%%%%%%%%%%%%%%%%%%%%%%%%%%%%%%
\begin{align}
-\Vamp^{\,\mathrm{sl},\star} &=
    \frac{1}{\vep^2}\bigg(\frac{\mu^2}{-s_{23}}\bigg)^\vep
    +\frac{3}{2\vep}\bigg(\frac{\mu^2}{-s_{23}}\bigg)^\vep
    +4
~,
\end{align}
%%%%%%%%%%%%%%%%%%%%%%%%%%%%%%%%%%%%%%%%
%
where we included the sub-leading color part of (\ref{Eq:deltaR}) and subtracted the
necessary sub-leading color terms following \Equation{oneloopauxq}.
For the finite contribution, we find
\begin{align}
\Famp^{\,\mathrm{sl},\star} &=
  \bigg[{-}\frac{  \SQR{p3}^2 \ANG{24}^2 }{  \kapp \ANG{45} s_{k3} \ANGSQR{2|k|3}  }
         +\frac{  \ANGSQR{3|k|5}^2 \ANGSQR{p|(4+5)|2}^2 }{  \kstr \SQR{45} s_{k3} \ANGSQR{3|k|2}^3 } \bigg] 
   \mathrm{Ls}_{-1}^{2mh}(s_{23},s_{k3},\kT^2,s_{45})
\notag\\&\hspace{0ex} 
 +\bigg[
     \frac{1}{2}\frac{ (3 s_{23} \delta_{23}-\Delta) (s_{k2}-s_{k3})(s_{p3}+s_{p2})\ANGSQR{4|k|5} }{ \ANGSQR{3|k|2} \Delta^2  }
   + \frac{1}{2}\frac{ s_{23} (s_{k2} - s_{k3} ) \ANG{4p} \SQR{p5}  }{ \ANGSQR{3|k|2} \Delta  } 
\notag\\&\hspace{4ex}
   + \frac{ \kapp \ANG{45} \ANGSQR{p|(2+3)|5}^2 }{ \ANGSQR{3|k|2} \Delta } 
   + \frac{ \ANG{p3} \SQR{25} \delta_{23} ( \kapp \ANG{4p}\delta_{k^2}-\ANGSQR{4|5|p}\delta_{45}) }{ \ANGSQR{3|k|2}^2 \Delta }
\notag\\&\hspace{4ex}
   - \frac{ \ANG{p3} \SQR{25} \ANGSQR{4|(2+5)|p}  }{ \ANGSQR{3|k|2}^2 }
   + 2 \frac{ \SQR{p3} \ANG{24} \ANGSQR{p|(4+5)|2} \ANGSQR{3|k|5}  }{ s_{k3}^2 \ANGSQR{3|k|2} } 
  \bigg]I_3(\kT^2,s_{23},s_{45})
\notag\\&\hspace{0ex}
   {-}\frac{ 1 }{ 2 } \frac{ \SQR{52}^2 \ANG{2p}^2 s_{k3} }{  \kstr \SQR{45} \ANGSQR{3|k|2} }
      \frac{ \mathrm{L}_1\big(\frac{ -s_{45} }{ -s_{k3} } \big) }{ s_{k3}^2 }
   -\frac{ 1 }{ 2 } \frac{ \SQR{p2}^2 \ANG{24}^2 s_{k2} }{  \kapp \ANG{54} \ANGSQR{3|k|2} }
       \frac{ \mathrm{L}_{1}\big(\frac{ -\kT^2 }{ -s_{k2} } \big) }{ s_{k2}^2 }
\notag\\&\hspace{0ex} 
    +\frac{\ANG{p2}\SQR{25}}{\kstr \SQR{45} \ANGSQR{3|k|2}}
     \bigg[2\ANGSQR{p|(2+4)|5}
           +\frac{ \ANG{p3} \SQR{25} s_{k3} }{  \ANGSQR{3|k|2} } 
     \bigg]\frac{ \mathrm{L}_0\big(\frac{ -s_{k3} }{ -s_{45} }\big) }{ s_{45}  }
\notag\\&\hspace{0ex}
   -\frac{ \ANG{42}\SQR{2p} }{\kapp \ANG{54} \ANGSQR{3|k|2}}
    \bigg[2\ANGSQR{4|(5+3)|p}
          +\frac{ \ANG{43} \SQR{2p} s_{k2} }{\ANGSQR{3|k|2}}
    \bigg]\frac{ \mathrm{L}_{0}\big(\frac{ -s_{k2} }{ -\kT^2 } \big) }{ \kT^2  }
\notag\\&\hspace{0ex}
     +\frac{\ANGSQR{p|(2+4)|5}  }{ \kstr \SQR{45} \ANGSQR{3|k|2} }
      \bigg( \frac{ 3 }{ 4 } \frac{ \ANGSQR{p|(2+4)|5} }{s_{k3}} 
            -\frac{ 1 }{ 2 } \frac{ \ANG{p3} \SQR{52} }{ \ANGSQR{3|k|2} }
      \bigg)\ln\bigg(\frac{ (-s_{k3}) (-s_{23}) }{ (-s_{45})^2 }\bigg)
\notag\\&\hspace{0ex}
     +\frac{\ANGSQR{4|(5+3)|p}}{ \kapp \ANG{54}\ANGSQR{3|k|2} }
      \bigg( \frac{ 3 }{ 4 } \frac{ \ANGSQR{4|(5+3)|p} }{s_{k2}}
            -\frac{ 1 }{ 2 } \frac{ \ANG{43} \SQR{p2}}{\ANGSQR{3|k|2}}
      \bigg)\ln\bigg(\frac{ (-s_{k2}) (-s_{23}) }{ (-\kT^2)^2 }\bigg) 
\notag\\&\hspace{0ex} 
     +\bigg[
         \frac{ 3 }{ 2 } \frac{\delta_{45} (s_{k2}-s_{k3})(s_{p3}+s_{p2})\ANGSQR{4|k|5}  }{ \ANGSQR{3|k|2}\Delta } 
       - \frac{ \kapp \ANG{p3} \SQR{25} }{ \ANGSQR{3|k|2}^2\Delta }\big(\ANG{p4} (s_{k2}-s_{k3})+2 \kstr \ANG{45}\SQR{5p} \big)
\notag\\&\hspace{4ex}
       +\frac{ \kapp \ANG{p4} \ANGSQR{p|(2-3)|5} }{ \ANGSQR{3|k|2} \Delta } 
       +\frac{ \SQR{p5} }{ \SQR{45} \ANGSQR{3|k|2} \Delta } 
          \bigg(\ANGSQR{p|3|5} s_{k3}-\ANGSQR{p|2|5} s_{k2}+\frac{ \ANG{p3} \SQR{25} \delta_{23} s_{k3} }{  \ANGSQR{3|k|2} } \bigg) 
      \bigg] 
\notag\\&\hspace{2ex}
   \times\ln\bigg(\frac{ -\kT^2  }{ -s_{23}  }\bigg) 
\notag\\&\hspace{0ex}
 +\bigg[
     \frac{ 3 }{ 2 } \frac{ \delta_{k^2} (s_{k2}-s_{k3})(s_{p2}+s_{p3}) \ANGSQR{4|k|5}}{ \ANGSQR{3|k|2} \Delta^2 }
   - \frac{ \SQR{54} \ANG{43} \SQR{2p} }{ \ANGSQR{3|k|2}^2 \Delta }\big(\ANG{4p} (s_{k2}-s_{k3})+2\kstr \ANG{45} \SQR{5p}\big)
\notag\\&\hspace{4ex} 
   -\frac{ \SQR{54} \ANG{p4} \ANGSQR{4|(2-3)|p} }{ \ANGSQR{3|k|2} \Delta } 
   -\frac{ \SQR{5p} }{ \kapp \ANGSQR{3|k|2} \Delta }\bigg(\ANGSQR{4|3|p} s_{k2}-\ANGSQR{4|2|p} s_{k3} -\frac{ \ANG{43} \SQR{2p} \delta_{23} s_{k2} }{  \ANGSQR{3|k|2} }\bigg)
  \bigg]
\notag\\&\hspace{2ex}
 \times\ln\bigg(\frac{ -s_{45}  }{ -s_{23}  }\bigg) 
\notag\\&\hspace{0ex} 
 + \frac{ 1 }{ 2 } \frac{ \SQR{p5} (s_{k3}-s_{k2}) (\SQR{p5} \delta_{23}+2 \kapp  \ANG{p4}\SQR{45}) }{ \kapp \SQR{45} \ANGSQR{3|k|2} \Delta }  
\;\;+\;\;\mathrm{flip}
~,
\end{align}
where $\mathrm{flip}$ again acts on all preceding terms.
Furthermore, we denote
%
%%%%%%%%%%%%%%%%%%%%%%%%%%%%%%%%%%%%%%%%
\begin{equation}
\delta_{k^2} = \kT^2 - s_{23} - s_{45}
\quad,\quad
\delta_{23} = s_{23} - s_{45} - \kT^2
\quad,\quad
\delta_{45} = s_{45} - \kT^2 - s_{23}
~,
\end{equation}
%%%%%%%%%%%%%%%%%%%%%%%%%%%%%%%%%%%%%%%%
%
and
%
%%%%%%%%%%%%%%%%%%%%%%%%%%%%%%%%%%%%%%%%
\begin{equation}
\Delta = (\kT^2)^2 + s_{23}^2 + s_{45}^2 - 2\kT^2s_{23}-2s_{23}s_{45}-2s_{45}\kT^2
~.
\end{equation}
%%%%%%%%%%%%%%%%%%%%%%%%%%%%%%%%%%%%%%%%
%
Also for the terms proportional to $1/\kapp$ in ${-}\Vamp^{\,\mathrm{sl},\star}\Amp^{\tree,\star}+\Famp^{\,\mathrm{sl},\star}$ we find the correct on-shell limit of $|\kT|\to0$, matching the expressions in Appendix~IV of~\cite{Bern:1997sc}.
The amplitude $\Amp_{3}$ in \Equation{Eq:qqee01} is the axial vector quark triangle contribution, and given by
%
%%%%%%%%%%%%%%%%%%%%%%%%%%%%%%%%%%%%%%%%
\begin{align}
&\Amp_{3}(1^\star,2_{\qb},3_{\qq},4_{\ePLS},5_{\eMIN})
\notag\\&\hspace{4ex}=
  \frac{2}{(4\pi)^2}\frac{f(m_t;s_{23},\kT^2,s_{45})-f(m_b;s_{23},\kT^2,s_{45})}{s_{45}}
  \bigg(
     \frac{\SQR{5p}\ANG{p2}\ANG{24}}{\ANG{23}}
   - \frac{\SQR{53}\SQR{3p}\ANG{p4}}{\SQR{23}}
  \bigg)
\notag\\&\hspace{4ex}
  + \frac{2}{(4\pi)^2}\frac{f(m_t;\kT^2,s_{23},s_{45})-f(m_b;\kT^2,s_{23},s_{45})}{s_{45}}
  \bigg(
     \frac{\SQR{53}\ANG{2p}\ANG{p4}}{\kstr}
   - \frac{\SQR{5p}\SQR{p3}\ANG{24}}{\kapp}
  \bigg)
~,
\label{Eq:qqee02}
\end{align}
%%%%%%%%%%%%%%%%%%%%%%%%%%%%%%%%%%%%%%%%
%
with
%
%%%%%%%%%%%%%%%%%%%%%%%%%%%%%%%%%%%%%%%%
\begin{align}
f(m;s,t,u) = \int_0^1da\int_0^1db\int_0^1dc\,\delta(1-a-b-c)\,\frac{bc}{m^2-sab-tbc-uca}
~.
\end{align}
%%%%%%%%%%%%%%%%%%%%%%%%%%%%%%%%%%%%%%%% 
%
This function simplifies for $m=0$, and can be expanded in $1/m^2$ for $m\to\infty$.
The expressions can be found in~\cite{Bern:1997sc}.
We do mention that when $|\kT|\to0$, only the last line in \Equation{Eq:qqee02} contributes, and we have
%
%%%%%%%%%%%%%%%%%%%%%%%%%%%%%%%%%%%%%%%%
\begin{equation}
f(m\to\infty;0,s_{23},s_{45})-f(0;0,s_{23},s_{45})
=
-\frac{1}{2s_{45}}\,\mathrm{L}_1\bigg(\frac{-s_{23}}{-s_{45}}\bigg) + \frac{1}{24m^2} + \Ord\big(1/m^4\big)
~.
\end{equation}
%%%%%%%%%%%%%%%%%%%%%%%%%%%%%%%%%%%%%%%%
%

\section{Conclusions}
In this paper, we explicitly calculate the virtual contributions for the one-jet processes $\procgg,\procqq$, $\procH$ and  $\procee$, and we check their collinear limit.
The off-shell scattering amplitudes for these processes have been systematically extracted from the on-shell embedding processes available in the literature according to the auxiliary parton method.
The main result of this work, besides providing  the explicit expressions the off-shell scattering amplitudes corresponding to the mentioned processes, is to prove, process by process, the conjecture developed in~\cite{vanHameren:2022mtk} for the structure of the NLO virtual contribution in hybrid $k_T$-factorization.
Our explicit calculations of the NLO virtual contributions to the parton level cross-sections of the mentioned processes completely agree with the conjecture formulated in Ref.~\cite{vanHameren:2022mtk}.
Furthermore, the explicit results obtained here will be useful to the theoretical community for the cross-section calculations at NLO within hybrid $k_T$-factorization scheme and will play an important role toward the establishment of a higher precision description of small $x$ phenomena. %
Our explicit calculations of the virtual corrections to the $\procgg,\procqq,\procH$ and $\procee$ LO cross sections show how to systematically derive the one loop amplitudes  from the on-shell helicity amplitudes. This is the core of our investigation: to show with explicit examples that our method can be successfully extended to NLO and to provide the guiding principles  to extract the virtual corrections  from the on-shell  amplitudes.
The computation of the real emission corrections still remains to be done and it is the subject of our present investigation.
Our  calculation scheme for the virtual corrections, which has been presented in this work, together with a general framework for computing the real corrections, which is under investigation, is an essential step forward toward the implementation of a numerical code for an event simulation generator, based on $k_T$ factorization, which can handle NLO calculations and which is therefore  able to provide a significant contribution in the rich field of small-x physics.

\subsection*{Acknowledgments}
EB acknowledge partial support by NCN grant No. DEC-2017/27/B/ST2/01985.
AG and AvH are supported by grant no.\ 2019/35/B/ST2/03531 of the Polish National Science Centre. PK is supported by the Polish National Science Centre, grant no. 2020/39/O/ST2/03011.

%

%%%%%%%%%%%%%%%%%%%%%%%%%%%%%%%%%%%%%%%%%%%%%%%%%%%%%%%%%%%%%%%%%%%%%%%%%%%%%%%%
%%% References %%%%%%%%%%%%%%%%%%%%%%%%%%%%%%%%%%%%%%%%%%%%%%%%%%%%%%%%%%%%%%%%%
%%%%%%%%%%%%%%%%%%%%%%%%%%%%%%%%%%%%%%%%%%%%%%%%%%%%%%%%%%%%%%%%%%%%%%%%%%%%%%%%
\bibliography{ls-refs}{}\bibliographystyle{JHEP}

\begin{appendix}
\addtocontents{toc}{\protect\setcounter{tocdepth}{1}}
%%%%%%%%%%%%%%%%%%%%%%%%%%%%%%%%%%%%%%%%%%%%%%%%%%%%%%%%%%%%%%%%%%%%%%%%%%%%%%%%
%%% Appendix: Special functions %%%%%%%%%%%%%%%%%%%%%%%%%%%%%%%%%%%%%%%%%%%%%%%%
%%%%%%%%%%%%%%%%%%%%%%%%%%%%%%%%%%%%%%%%%%%%%%%%%%%%%%%%%%%%%%%%%%%%%%%%%%%%%%%%
\section{Special functions}
We use the same functions as defined in for example~\cite{Bern:1997sc}.
There is
%
%%%%%%%%%%%%%%%%%%%%%%%%%%%%%%%%%%%%%%%%
\begin{equation}
\mathrm{Li}_2(x) = -\int_0^xdz\,\frac{\ln(1-z)}{z}
~,
\end{equation}
%%%%%%%%%%%%%%%%%%%%%%%%%%%%%%%%%%%%%%%%
%
which satisfies the useful relation
%
%%%%%%%%%%%%%%%%%%%%%%%%%%%%%%%%%%%%%%%%
\begin{equation}
\mathrm{Li}_2\big(1-x^{-1}\big) = - \mathrm{Li}_2(1-x) - \frac{1}{2}\ln^2(x)
~.
\end{equation}
%%%%%%%%%%%%%%%%%%%%%%%%%%%%%%%%%%%%%%%%
%
Then there are
%
%%%%%%%%%%%%%%%%%%%%%%%%%%%%%%%%%%%%%%%%
\begin{equation}
\mathrm{L}_0(x) = \frac{\ln(x)}{1-x}
\quad,\quad
\mathrm{L}_1(x) = \frac{\ln(x)+1-x}{(1-x)^2}
~,
\end{equation}
%%%%%%%%%%%%%%%%%%%%%%%%%%%%%%%%%%%%%%%%
%
which satisfy the useful relations
%
%%%%%%%%%%%%%%%%%%%%%%%%%%%%%%%%%%%%%%%%
\begin{equation}
\mathrm{L}_0\big(x^{-1}\big) = x\,\mathrm{L}_0(x)
\quad,\quad
\mathrm{L}_1\big(x^{-1}\big) = -x^2\,\mathrm{L}_1(x) - x
~.
\end{equation}
%%%%%%%%%%%%%%%%%%%%%%%%%%%%%%%%%%%%%%%%
%
The function
%
%%%%%%%%%%%%%%%%%%%%%%%%%%%%%%%%%%%%%%%%
\begin{align}
\mathrm{Ls}_{-1}(x,y) &= \mathrm{Li}_2(1-x) + \mathrm{Li}_2(1-y) + \ln(x)\ln(y) -\frac{\pi^2}{6}
\end{align}
%%%%%%%%%%%%%%%%%%%%%%%%%%%%%%%%%%%%%%%%
%
does not show up in any of our expressions, but will appear in the limit of a function we do use and present below.
First, however, we present
%
%%%%%%%%%%%%%%%%%%%%%%%%%%%%%%%%%%%%%%%%
\begin{equation}
\mathrm{I}_3^{3m}(x,y,z) = 
   \int_0^1da\int_0^1db\int_0^1dc\,\delta(1-a-b-c)\,\frac{-1}{xab+ybc+zca}
~,
\end{equation}
%%%%%%%%%%%%%%%%%%%%%%%%%%%%%%%%%%%%%%%%
%
which is symmetric in its arguments, and satisfies the limit
%
%%%%%%%%%%%%%%%%%%%%%%%%%%%%%%%%%%%%%%%%
\begin{equation}
\mathrm{I}_3^{3m}(x,y,z\to0) =
 \frac{1}{x-y}\bigg[
                    \ln\frac{x}{y}\ln\frac{z}{y}
                    +2\mathrm{Li}_2\bigg(1-\frac{x}{y}\bigg)
              \bigg]
 +\Ord(z)
~.
\end{equation}
%%%%%%%%%%%%%%%%%%%%%%%%%%%%%%%%%%%%%%%%
%
And finally, we use
%
%%%%%%%%%%%%%%%%%%%%%%%%%%%%%%%%%%%%%%%%
\begin{align}
\mathrm{Ls}_{-1}^{2mh}(x_1,x_2,y_1,y_2) &= 
   -\mathrm{Li}_2\bigg(1-\frac{y_1}{x_2}\bigg)
   -\mathrm{Li}_2\bigg(1-\frac{y_2}{x_2}\bigg)
   -\frac{1}{2}\ln^2\bigg(\frac{-x_1}{-x_2}\bigg)
\\&\hspace{0ex}
   +\frac{1}{2}\ln\bigg(\frac{-x_1}{-y_1}\bigg)\ln\bigg(\frac{-x_1}{-y_2}\bigg)
   +\frac{1}{2}\bigg[\frac{x_1-y_1-y_2}{2}+\frac{y_1y_2}{x_2}\bigg]\mathrm{I}_3^{3m}(x_1,y_1,y_2)
~,
\notag
\end{align}
%%%%%%%%%%%%%%%%%%%%%%%%%%%%%%%%%%%%%%%%
%
which satisfies the limit
%
%%%%%%%%%%%%%%%%%%%%%%%%%%%%%%%%%%%%%%%%
\begin{equation}
\mathrm{Ls}_{-1}^{2mh}(x_1,x_2,y,0) = 
   \mathrm{Ls}_{-1}\bigg(\frac{x_1}{y},\frac{x_2}{y}\bigg)
~.
\end{equation}
%%%%%%%%%%%%%%%%%%%%%%%%%%%%%%%%%%%%%%%%
%

\section{\label{App:ggauxg}$\BLDempty\BLDto\BLDgSTR\BLDg\BLDg$ from auxiliary gluons}
Given the general color decomposition of \Equation{Eq:def09} and the fact that partial amplitudes with non-adjacent auxiliary gluons do not contribute, the tree-level decomposition can be written in a compact notation as
%
%%%%%%%%%%%%%%%%%%%%%%%%%%%%%%%%%%%%%%%%
\begin{equation}
\fullAmp^{\tree}(\auxA,2,3,\auxB) 
    = T_{\auxA 23\auxB}\Amp^{\tree}_{\auxA 23\auxB}
    + T_{\auxA 32\auxB}\Amp^{\tree}_{\auxA 32\auxB}
    + T_{\auxB 23\auxA}\Amp^{\tree}_{\auxB 23\auxA}
    + T_{\auxB 32\auxA}\Amp^{\tree}_{\auxB 32\auxA}
~,
\end{equation}
%%%%%%%%%%%%%%%%%%%%%%%%%%%%%%%%%%%%%%%%
%
where
%
%%%%%%%%%%%%%%%%%%%%%%%%%%%%%%%%%%%%%%%%
\begin{equation}
T_{ijkl} = \Tr\big(T^{a_{i}}T^{a_{j}}T^{a_{k}}T^{a_{l}}\big)
\quad\textrm{and}\quad
\Amp^{\tree}_{ijkl} = \AmpTree(i,j,k,l)
~.
\end{equation}
%%%%%%%%%%%%%%%%%%%%%%%%%%%%%%%%%%%%%%%%
%
This can be reorganized as
%
%%%%%%%%%%%%%%%%%%%%%%%%%%%%%%%%%%%%%%%%
\begin{align}
\fullAmp^{\tree}(\auxA,2,3,\auxB) 
 &= \big(T_{\auxA 23\auxB}-T_{\auxA 32\auxB}-T_{\auxB 23\auxA}+T_{\auxB 32\auxA}\big)\Amp^{\tree}_{\auxA 23\auxB}
  + T_{\auxA 32\auxB}\big(\Amp^{\tree}_{\auxA 23\auxB}+\Amp^{\tree}_{\auxA 32\auxB}\big)
\notag\\&\hspace{16ex}
  + T_{\auxB 23\auxA}\big(\Amp^{\tree}_{\auxA 23\auxB}+\Amp^{\tree}_{\auxB 23\auxA}\big)
  + T_{\auxB 32\auxA}\big(\Amp^{\tree}_{\auxB 32\auxA}-\Amp^{\tree}_{\auxA 23\auxB}\big)
\notag\\&
=\big(\sqrt{2}\,\imag f^{\auxB\auxA c}\big)\big(\sqrt{2}\,\imag f^{23c}\big)\Amp^{\tree}_{\auxA 23\auxB}
  + T_{\auxA 32\auxB}\big(\Amp^{\tree}_{\auxA 23\auxB}+\Amp^{\tree}_{\auxA 32\auxB}\big)
\notag\\&\hspace{16ex}
  + T_{\auxB 23\auxA}\big(\Amp^{\tree}_{\auxA 23\auxB}+\Amp^{\tree}_{\auxB 23\auxA}\big)
  + T_{\auxB 32\auxA}\big(\Amp^{\tree}_{\auxB 32\auxA}-\Amp^{\tree}_{\auxA 23\auxB}\big)
~.
\end{align}
%%%%%%%%%%%%%%%%%%%%%%%%%%%%%%%%%%%%%%%%
%
The pairs of tree-level amplitudes appearing with the same trace structure vanish in the $\Lambda$ limit, and we are left with the same result as with auxiliary quarks.
At one loop we have, following \Equation{Eq:def08} and again including only amplitudes for which the auxiliary partons are adjacent, the color decomposition
%
%%%%%%%%%%%%%%%%%%%%%%%%%%%%%%%%%%%%%%%%
\begin{align}
\fullAmp(\auxA,2,3,\auxB) 
 &= \Nc\big(T_{\auxA 23\auxB}-T_{\auxA 32\auxB}-T_{\auxB 23\auxA}+T_{\auxB 32\auxA}\big)\Amp_{\auxA 23\auxB}
  + \Nc T_{\auxA 32\auxB}\big(\Amp_{\auxA 23\auxB}+\Amp_{\auxA 32\auxB}\big)
\notag\\&\hspace{8ex}
  + \Nc T_{\auxB 23\auxA}\big(\Amp_{\auxA 23\auxB}+\Amp_{\auxB 23\auxA}\big)
  + \Nc T_{\auxB 32\auxA}\big(\Amp_{\auxB 32\auxA}-\Amp_{\auxA 23\auxB}\big)
\notag\\&\hspace{16ex}
  + T_{\auxA\auxB}T_{23}\Amp_{\auxA\auxB;23}
  + T_{\auxA2}T_{\auxB3}\Amp_{\auxA2;\auxB3}
  + T_{\auxA3}T_{2\auxB}\Amp_{\auxA3;2\auxB}
~,
\end{align}
%%%%%%%%%%%%%%%%%%%%%%%%%%%%%%%%%%%%%%%%
%
with
%
%%%%%%%%%%%%%%%%%%%%%%%%%%%%%%%%%%%%%%%%
\begin{equation}
T_{ij} = \Tr\big(T^{a_{i}}T^{a_{j}}\big)=\delta^{a_{i}a_{j}}
\quad\textrm{and}\quad
\Amp_{ijkl} = \Amp_1(i,j,k,l)
\quad\textrm{and}\quad
\Amp_{ij;kl} = \Amp_3(i,j,k,l)
~.
\end{equation}
%%%%%%%%%%%%%%%%%%%%%%%%%%%%%%%%%%%%%%%%
%
The amplitudes $\Amp_3$ can be expressed in terms of $\Amp_1$ following Eq.~(6.6) in~\cite{Bern:1990ux}, and including only the ones that survive the $\Lambda$-limit, we have
%
%%%%%%%%%%%%%%%%%%%%%%%%%%%%%%%%%%%%%%%%
\begin{equation}
\Amp_{\auxA\auxB;23} = \Amp_{\auxA2;\auxB3} = \Amp_{\auxA3;2\auxB}
= \Amp_{A23B} +\Amp_{A32B} +\Amp_{B23A} +\Amp_{B32A}
~.
\end{equation}
%%%%%%%%%%%%%%%%%%%%%%%%%%%%%%%%%%%%%%%%
%
Notice that this would vanish at tree level, as expected.
At one loop, the combinations do not quite vanish, and we find
%
%%%%%%%%%%%%%%%%%%%%%%%%%%%%%%%%%%%%%%%%
\begin{align}
 \Amp_{\auxA 23\auxB}+\Amp_{\auxA 32\auxB}
     & \lamlim -2\mysign(s_{p2})\AmpRt(1^\star,2,3)
\\
 \Amp_{\auxA 23\auxB}+\Amp_{\auxB 23\auxA}
     & \lamlim -2\mysign(s_{p2})\AmpRt(1^\star,2,3)
\\
 \Amp_{\auxB 32\auxA}-\Amp_{\auxA 23\auxB}
     & \lamlim 0
~.
\end{align}
%%%%%%%%%%%%%%%%%%%%%%%%%%%%%%%%%%%%%%%%
%
Furthermore, we find
%
%%%%%%%%%%%%%%%%%%%%%%%%%%%%%%%%%%%%%%%%
\begin{equation}
\Amp_{\auxA 23\auxB} \lamlim \Amp^{\reLab}(1^\star,2,3)-\mysign(s_{p2})\AmpRt(1^\star,2,3)
\end{equation}
%%%%%%%%%%%%%%%%%%%%%%%%%%%%%%%%%%%%%%%%
%
with $\Amp^{\reLab}$ from \Equation{Eq:gggre}, and thus
%
%%%%%%%%%%%%%%%%%%%%%%%%%%%%%%%%%%%%%%%%
\begin{align}
\fullAmp(\auxA,2,3,\auxB) &\lamlim
   \Nc\big(T_{\auxA 23\auxB}-T_{\auxA 32\auxB}-T_{\auxB 23\auxA}+T_{\auxB 32\auxA}\big)\Amp^{\reLab}(1^\star,2,3)
\notag\\&\hspace{8ex}
- \Big[
  \Nc\big(T_{\auxA 23\auxB}-T_{\auxA 32\auxB}-T_{\auxB 23\auxA}+T_{\auxB 32\auxA}\big)
  +2\Nc\big(T_{\auxA 32\auxB}+T_{\auxB 23\auxA}\big)
\notag\\&\hspace{20ex}
  +4\big(T_{\auxA\auxB}T_{23}
       + T_{\auxA2}T_{\auxB3}
       + T_{\auxA3}T_{2\auxB}\big)
  \Big]\mysign(s_{p2})\AmpRt(1^\star,2,3)
\notag\\&\hspace{2ex}
=
\big(\sqrt{2}\,\imag f^{\auxB\auxA c}\big)\big(\sqrt{2}\,\imag f^{23c}\big)\Nc\Amp^{\reLab}(1^\star,2,3)
\notag\\&\hspace{8ex}
- \Big[
 \Nc\big(T_{\auxA 23\auxB}+T_{\auxA 32\auxB}+T_{\auxB 23\auxA}+T_{\auxB 32\auxA}\big)
\notag\\&\hspace{16ex}
  +4\big(T_{\auxA\auxB}T_{23}
       + T_{\auxA2}T_{\auxB3}
       + T_{\auxA3}T_{2\auxB}\big) 
\Big]\mysign(s_{p2})\AmpRt(1^\star,2,3)
~.
\end{align}
%%%%%%%%%%%%%%%%%%%%%%%%%%%%%%%%%%%%%%%%
%
The last two lines are symmetric in $\auxA,\auxB$ and $2,3$, and do not contribute at NLO.

\section{\label{App:qqauxg}$\BLDempty\BLDto\BLDgSTR\BLDqB\BLDq$ from auxiliary quarks}
From~\cite{Kunszt:1993sd}, we have the formula
%
%%%%%%%%%%%%%%%%%%%%%%%%%%%%%%%%%%%%%%%%
\begin{align}
\fullAmp(2_{\qb},\auxB_{\QB},\auxA_{\QQ},3_{\qq})
&=
\bigg(\delta^{\imath_{\auxB}}_{i_{3}}\delta^{\imath_{2}}_{i_{\auxA}}
        -\frac{1}{\Nc}\delta^{\imath_{2}}_{i_{3}}\delta^{\imath_{\auxB}}_{i_{\auxA}}\bigg)a_1(2,\auxB;\auxA,3)
+ \delta^{\imath_{\auxB}}_{i_{3}}\delta^{\imath_{2}}_{i_{\auxA}}\,a_2(2,\auxB;\auxA,3)
\notag\\&=
\bigg(\delta^{\imath_{\auxB}}_{i_{3}}\delta^{\imath_{2}}_{i_{\auxA}}
        -\frac{1}{\Nc}\delta^{\imath_{2}}_{i_{3}}\delta^{\imath_{\auxB}}_{i_{\auxA}}\bigg)\big[a_1(2,\auxB;\auxA,3)+a_2(2,\auxB;\auxA,3)\big]
\\&\hspace{12ex}
+ \frac{1}{\Nc}\delta^{\imath_{2}}_{i_{3}}\delta^{\imath_{\auxB}}_{i_{\auxA}}\,a_2(2,\auxB;\auxA,3)
\notag~,
\end{align}
%%%%%%%%%%%%%%%%%%%%%%%%%%%%%%%%%%%%%%%%
%
where the last line vanishes upon contraction with the tree-level color structure.
For the partial amplitudes, we find, after ``un-doing'' the UV subtraction by putting $\beta_0=0$,
%
%%%%%%%%%%%%%%%%%%%%%%%%%%%%%%%%%%%%%%%%
\begin{equation}
a_1(2,\auxB;\auxA,3)+a_2(2,\auxB;\auxA,3) \lamlim
\Amp^{\reLab}(1^\star,2_{\qb},3_{\qq})
-\bigg(1-\frac{4}{\Nc^2}\bigg)\mysign(s_{p2})\AmpRt(1^\star,2_{\qb},3_{\qq})
~,
\end{equation}
%%%%%%%%%%%%%%%%%%%%%%%%%%%%%%%%%%%%%%%%
%
with $\Amp^{\reLab}$ from \Equation{Eq:gqqre}.
%
%{\color{red}%
Again, with the same reasoning as argued there, we see that the contributions proportional to $\mysign$ vanish in the combination of \Equation{symauxq}.
%
%}%\color{red}

\section{\label{App:eeauxg}$\BLDempty\BLDto\BLDgSTR\BLDqB\BLDq\BLDePLS\BLDeMIN$ from auxiliary gluons}
We omit the arguments $(4_{e^+},5_{e^-})$ from all amplitudes here.
The tree-level partial amplitudes satisfy
%%%%%%%%%%%%%%%%%%%%%%%%%%%%%%%%%%%%%%%%
\begin{align}
\Amp^{\tree}(3_{q},\auxA,\auxB,2_{\bar{q}}) &\lamlim \Amp^{\tree}(1^\star,2_{\bar{q}},3_{q})
~,\\
\Amp^{\tree}(3_{q},\auxB,\auxA,2_{\bar{q}}) &\lamlim -\Amp^{\tree}(1^\star,2_{\bar{q}},3_{q})
~,
\end{align}
%%%%%%%%%%%%%%%%%%%%%%%%%%%%%%%%%%%%%%%%
%
so, from Eq.~(2.6) in~\cite{Bern:1997sc} it is clear that the tree-level amplitude is proportional to $f^{a_{\auxA}a_{\auxB}a_1}\big(T^{a_1}\big)_{i_{3}}^{\imath_{2}}$.
Thus, the only terms of the one-loop amplitude of Eq.~(2.9) in~\cite{Bern:1997sc} that are relevant and survive contraction with the tree-level color structure are
%
%%%%%%%%%%%%%%%%%%%%%%%%%%%%%%%%%%%%%%%%
\begin{align}
\fullAmp^{\mathrm{relevant}}(3_{q},\auxA,\auxB,2_{\bar{q}})
&=
\Cone\Nc\Big\{
\big(T^{a_{\auxA}}T^{a_{\auxB}}\big)_{i_{3}}^{\imath_{2}}\Amp_1(3_{q},\auxA,\auxB,2_{\bar{q}})
+
\big(T^{a_{\auxB}}T^{a_{\auxA}}\big)_{i_{3}}^{\imath_{2}}\Amp_1(3_{q},\auxB,\auxA,2_{\bar{q}})
\Big\}
\notag\\&\hspace{-4ex}
+ \Cthree\Big\{
\big(T^{a_{\auxA}}T^{a_{\auxB}}\big)_{i_{3}}^{\imath_{2}}\Amp^{\mathrm{ax}}_4(3_{q},2_{\bar{q}};\auxA,\auxB)
+
\big(T^{a_{\auxB}}T^{a_{\auxA}}\big)_{i_{3}}^{\imath_{2}}\Amp^{\mathrm{ax}}_4(3_{q},2_{\bar{q}};\auxB,\auxA)
\Big\}
~.
\end{align}
%%%%%%%%%%%%%%%%%%%%%%%%%%%%%%%%%%%%%%%%
%
This can be reorganized to
%
%%%%%%%%%%%%%%%%%%%%%%%%%%%%%%%%%%%%%%%%
\begin{align}
\fullAmp^{\mathrm{relevant}}(3_{q},\auxA,\auxB,2_{\bar{q}})
&=
 f^{a_{\auxA}a_{\auxB}a_1}\big(T^{a_1}\big)_{i_{3}}^{\imath_{2}}\bigg\{
   \Cone\Nc\Amp_1(3_{q},\auxA,\auxB,2_{\bar{q}})
      +\Cthree\Amp^{\mathrm{ax}}_4(3_{q},2_{\bar{q}};\auxA,\auxB)
\notag\\&\hspace{14ex}
  -\frac{\Cone\Nc}{2}\Big[
       \Amp_1(3_{q},\auxA,\auxB,2_{\bar{q}})
      +\Amp_1(3_{q},\auxB,\auxA,2_{\bar{q}})
   \Big]
\\&\hspace{16ex}
  -\frac{\Cthree}{2}\Big[
       \Amp^{\mathrm{ax}}_4(3_{q},2_{\bar{q}};\auxA,\auxB)
      +\Amp^{\mathrm{ax}}_4(3_{q},2_{\bar{q}};\auxB,\auxA)
   \Big]
\bigg\}
\notag~,
\end{align}
%%%%%%%%%%%%%%%%%%%%%%%%%%%%%%%%%%%%%%%%
%
where we omitted contributions proportional to $\big[\big(T^{a_{\auxA}}T^{a_{\auxB}}\big)_{i_{3}}^{\imath_{2}} +\big(T^{a_{\auxB}}T^{a_{\auxA}}\big)_{i_{3}}^{\imath_{2}}\big]$, because they vanish upon contraction with the tree-level color structure.
%

%{\color{red}%
Now, it is the second line that would cause a difference with the result obtained with auxiliary quarks.
It is the equivalent of the second line of \Equation{Eq:qqee05}, and one can immediately see that the factors $2/\Nc^2$ and $\Nc/2$ do not match.
This contribution above clearly becomes symmetric in $\auxA,\auxB$ if summed over the auxiliary parton helicities, and then vanishes in the combination of \Equation{symauxg}.
%}%\color{red}

%
The decomposition of $\Amp_1$ into primitive amplitudes from~\cite{Bern:1997sc} is
%
%%%%%%%%%%%%%%%%%%%%%%%%%%%%%%%%%%%%%%%%
\begin{align}
\Amp_1(3_{q},\auxA,\auxB,2_{\bar{q}})
&= \Amp(3_{q},\auxA,\auxB,2_{\bar{q}}) - \frac{1}{\Nc^2}\Amp(3_{q},2_{\bar{q}},\auxB,\auxA)
\notag\\&\hspace{17ex}
  - \frac{n_f}{\Nc}\big[\Amp^{f}(3_{q},\auxA,\auxB,2_{\bar{q}})+\Amp^{s}(3_{q},\auxA,\auxB,2_{\bar{q}})\big]
~,
\end{align}
%%%%%%%%%%%%%%%%%%%%%%%%%%%%%%%%%%%%%%%%
%
where we omitted the top-loop contributions that behave as $|\kT|^2/m_t^2$ again.
We confirm that, referring to the amplitudes in \Equation{Eq:qqee03},
%
%%%%%%%%%%%%%%%%%%%%%%%%%%%%%%%%%%%%%%%%
\begin{align}
\cGamma^{-1}&\Amp(3_{q}^+,\auxA^+,\auxB^-,2_{\bar{q}}^-) \lamlim 
   \Vamp^{\,\star}\Amp^{\tree,\star}
  +\Famp^{\,\mathrm{re},\star} 
  +\Theta(s_{p2})\Amp^{\mathrm{Rt},\star}  
  + \mysign(s_{p2})\Famp^{\,\mathrm{im},\star}
~,\\
\cGamma^{-1}&\Amp(3_{q}^+,2_{\bar{q}}^-,\auxB^-,\auxA^+) \lamlim 
\Vamp^{\,\mathrm{sl},\star} \Amp^{\tree,\star} -\Famp^{\,\mathrm{sl},\star}
~,\\
&\Amp^{f}(3_{q}^+,\auxA^+,\auxB^-,2_{\bar{q}}^-)+\Amp^{s}(3_{q}^+,\auxA^+,\auxB^-,2_{\bar{q}}^-) \lamlim  0
~.
\end{align}
%%%%%%%%%%%%%%%%%%%%%%%%%%%%%%%%%%%%%%%%
%
We also confirm
%
%%%%%%%%%%%%%%%%%%%%%%%%%%%%%%%%%%%%%%%%
\begin{equation}
\Amp_{4}^{\mathrm{ax}}(3_{q}^+,2_{\bar{q}}^-;\auxA^+,\auxB^-)
+
\Amp_{4}^{\mathrm{ax}}(3_{q}^+,2_{\bar{q}}^-,\auxB^-;\auxA^+)
\lamlim
0
~,
\end{equation}
%%%%%%%%%%%%%%%%%%%%%%%%%%%%%%%%%%%%%%%%
%
and we confirm
%
%%%%%%%%%%%%%%%%%%%%%%%%%%%%%%%%%%%%%%%%
\begin{equation}
\Amp_4^{\mathrm{ax}}(3_{q}^+,2_{\bar{q}}^-;\auxA^+,\auxB^-)
\lamlim
\Amp_3(1^\star,2_{\bar{q}}^-,3_{q}^+)
\end{equation}
%%%%%%%%%%%%%%%%%%%%%%%%%%%%%%%%%%%%%%%%
%
in the large $m_t$ and $m_b=0$ approximation.
The latter involves the expression for $C^{\mathrm{ax}}$ in Eq.~(11.8) of~\cite{Bern:1997sc}, which is the only one we encountered with individual terms that behave as $\Lambda^2$, while the whole expressions behaves as $\Lambda$.

%{\color{blue}
\section{\label{weylspinors}Spinor helicity method}
In presenting our results we use the well-known spinor helicity formalism, which is the basis language of modern scattering amplitude calculations.
Let $u(p)$ and $v(p)$ respectively denote the positive and negative energy solutions to the massless Dirac equation
\begin{equation}
\slashp u(p)=  \slashp v(p) = 0 \,,
\label{DiracMassless}
\end{equation}
where the slash notation stands for a contraction with the gamma matrices, $\slashp \equiv \gamma_{\mu} p^{\mu}$, the latter obeying to the Clifford algebra, $\{ \gamma_{\mu},\gamma_{\nu} \} = 2 g_{\mu \nu}$, and  $g_{\mu \nu}$ is the Minkowski metric with signature $(1,-1,-1,-1)$.
When dealing with massless spinors an explicit useful realization of the gamma matrices is the Weyl representation:
\begin{equation}
\gamma_{\mu}=  \left(\begin{array}{cc} 0 & \sigma_{\mu} \\
\bar{\sigma}_{\mu} & 0
 \end{array}\right),\;\;\;
 \gamma^5 = i \gamma^0 \gamma^1 \gamma^2 \gamma^3 = \left(\begin{array}{cc} -I_2 & 0 \\
0& I_2
 \end{array}\right)
\end{equation}
where $\sigma_{\mu}=(I_2,\vec{\sigma})$ and  $\bar{\sigma}_{\mu}=(I_2,-\vec{\sigma})$ and $\vec{\sigma}=(\sigma_1,\sigma_2,\sigma_3)$ are the standard Pauli matrices, defined as:
\begin{equation}
\sigma_1= \left(\begin{array}{cc} 0 & 1 \\
1 &  0
 \end{array}\right),\;\;\;\sigma_2= \left(\begin{array}{cc} 0 & -i \\
i &  0
 \end{array}\right),\;\;\;\sigma_3= \left(\begin{array}{cc} 1 & 0 \\
0 &  -1
 \end{array}\right)\;.
\end{equation}
We can use the matrix $\gamma_5$ to construct the projection operators onto the upper and lower parts the four-component Dirac spinors $u(p)$ and $v(p)$:
\begin{equation}
u_{\pm}(p)=\frac{1\pm \gamma_5}{2}u(p), \;\;\; \text{and}\;\;\;  v_{\pm}(p)=\frac{1 \mp \gamma_5}{2} v(p)\,,
\end{equation}
where $u_{\pm}(p)$ are and $v_{\pm}(p)$ are the solutions of the massless Dirac equation with definite helicity, which is the projection of the spin along the  particle momentum. 
%A similar relation holds between the complex conjugated spinors:
%\begin{equation}
%\bar{u}_{\pm}(k)=\bar{u}(p) \frac{1\mp \gamma_5}{2} =\bar{v}_{\mp}(k)=\bar{v}(p) \frac{1 \mp \gamma_5}{2}\;.
%\end{equation}
Since in the massless limit $u_{+}(p)=v_{-}(p)$ and $u_{-}(p)=v_{+}(p)$ it is useful to introduce the following short-hand notation:
\begin{equation}
|p] \equiv  u_{-}(p) =  \left(\begin{array}{c} L(p)  \\
0
 \end{array}\right)
 \;\; \text{and}\;\;
 |p\rangle \equiv  u_{+}(p) = \left(\begin{array}{c} 0  \\
R(p)
 \end{array}\right)
\end{equation}
where  $L(p)$ and $R(p)$ are found by imposing that $u_{+}(p) $ and $u_{-}(p)$ satisfy the massless Dirac equation, Eq. \ref{DiracMassless}, and they can be explicitly written in the Weyl representation of the gamma matrices as follows \cite{vanHameren:2014iua}:
\begin{equation}
L(p)=\frac{1}{\sqrt{|p_0+p_3|}} \left(\begin{array}{c} - p_1+ i p_2   \\
 p_0+  p_3 
 \end{array}\right), \;\;\;\;\;\;
 R(p)=\frac{\sqrt{|p_0+p_3|}}{p_0+p_3|} \left(\begin{array}{c}  p_0+  p_3   \\
 p_1 + i p_2 
 \end{array}\right)\,.
\end{equation}
The “dual” Weyl spinors are defined as \cite{vanHameren:2014iua}:
\begin{equation}
[p| = ((\varepsilon L(p))^T ,0),\;\;\; \langle p| = (0,(\varepsilon^T R(p))^T),\;\;\;\text{where} \;\;\; \varepsilon= \left(\begin{array}{cc} 0 & 1 \\
-1 &  0
 \end{array}\right)\,.
\end{equation}
%The algebra is simplified if we take  matrices in Weyl representation:
We use the following conventions for the spinor products:
\begin{equation}
\langle k  l \rangle = \bar{u}_{-}(k) u_{+}(l); \;\; [ k  l ] = \bar{u}_{+}(k) u_{-}(l)\,.
\end{equation}
Observe that the spinor products are, up to a phase, square roots of Lorentz products:
\begin{equation}
\langle k l \rangle = \sqrt{2 k l} e^{i \eta(k,l)}\,,\;\;\; [  l k ] = \sqrt{2 k l} e^{-i \eta(k,l)}
\end{equation}
which implies the very useful relation
\begin{equation}
\langle k l \rangle [ l k ]= 2 k l=s_{kl}\,.
\end{equation}
Note that the mapping between a light-like vector and a Weyl spinor is non-linear: the spinor $|p\rangle + |q\rangle$ is another massless spinor, but it cannot be used to represent the four-vector $p+q$, which is not light-like.
On the other hand we have the linear maps:
\begin{equation}
| p \rangle [p| = \frac{1+\gamma_5}{2} \slashp, \;\;\; \text{and}\;\;\; | p]  \langle p | = \frac{1-\gamma_5}{2} \slashp
\end{equation}
which implies another very useful relation
\begin{equation}
| p \rangle [p| + | p]  \langle p | = \slashp\,.
\end{equation}
The spinors products also satisfy the Gordon identity:
\begin{equation}
\langle i | \gamma_{\mu} | i ] = 2 p_{\mu},
\end{equation}
antisymmetry:
\begin{equation}
\langle i j \rangle = - \langle  j i \rangle, \;\;\; [ i j ] = - [  j i ], \;\;\; \langle i i \rangle  = [i i]=0,
\end{equation}
Fierz rearrangement:
\begin{equation}
\langle i | \gamma_{\mu} | j ]  \langle k | \gamma^{\mu} | l ] =  \langle i  k \rangle  [j l],
\end{equation}
and the Schouten identities:
\begin{equation}
\langle p q  \rangle  \langle k l \rangle = \langle p k \rangle \langle q l \rangle + \langle p l \rangle \langle k q \rangle,
\;\;\; [ p q  ] [ k l ] = [ p k ] [ q l ] + [ p l ] [ k q ]\,.
\end{equation}
Given two  light-like vectors, $p$ (momentum of a massless particle) and $n$ (any external light-like vector), the polarisation vector for a gluon with a given helicity,
$\epsilon^{\mu}_{\lambda}(p)$, and momentum $p$, can be expressed as spinor products as follows \cite{vanHameren:2014iua}:
\begin{equation}
\epsilon_{+}^{\mu}(p)= \frac{\langle n  | \gamma^{\mu} | p ] }{\sqrt{2} \langle n  p \rangle}, \;\;\; \epsilon_{-}^{\mu}(p)= \frac{\langle p  | \gamma^{\mu} | n ] }{\sqrt{2} [ p  n ]}
\end{equation}
A good choice of the auxiliary vectors $n$ can simplify the calculations of helicity amplitudes enormously.
%}

\end{appendix}

\end{document}